%% file: ms.tex
\documentclass[acmtog]{acmart}

\usepackage{booktabs}

\usepackage[ruled]{algorithm2e}

\SetAlFnt{\small}
\SetAlCapFnt{\small}
\SetAlCapNameFnt{\small}
\SetAlCapHSkip{0pt}

\usepackage{subcaption}
\usepackage{kbordermatrix}
\usepackage{microtype}
\usepackage{placeins}
\usepackage{multirow}
\usepackage[export]{adjustbox}
\usepackage{csquotes}

\usepackage{todonotes}
\usepackage{color}

\newcommand{\eg}{\emph{e.g.}\xspace}
\newcommand{\ie}{\emph{i.e.}\xspace}
\newcommand{\cf}{\emph{cf.}\xspace}

\newcommand{\rtb}[1]{\rotatebox[origin=c]{90}{#1}}

\graphicspath{{figures/}}

\acmJournal{TOG}

\begin{document}
	\title{AlSub: Fully Parallel and Modular Subdivision}

\author{Daniel Mlakar}
\affiliation
{
	\institution{Graz University of Technology}
	\streetaddress{Inffeldgasse 16/II}
	\city{Graz}
	\state{Austria}
	\postcode{8010}
}
\email{daniel.mlakar@icg.tugraz.at}

\author{Martin Winter}
\affiliation
{
	\institution{Graz University of Technology}
	\streetaddress{Inffeldgasse 16/II}
	\city{Graz}
	\state{Austria}
	\postcode{8010}
}
\email{martin.winter@icg.tugraz.at}

\author{Hans-Peter Seidel}
\affiliation
{
	\institution{Max Planck Institute for Informatics}
	\streetaddress{Campus E1 4, Stuhlsatzenhausweg}
	\city{Saarbr{\"u}cken}
	\state{Germany}
	\postcode{66123}
}
\email{hpseidel@mpi-inf.mpg.de}

\author{Markus Steinberger}
\affiliation
{
	\institution{Graz University of Technology}
	\streetaddress{Inffeldgasse 16/II}
	\city{Graz}
	\state{Austria}
	\postcode{8010}
}
\email{steinberger@icg.tugraz.at}	

\author{Rhaleb Zayer}
\affiliation
{
	\institution{Max Planck Institute for Informatics}
	\streetaddress{Campus E1 4, Stuhlsatzenhausweg}
	\city{Saarbr{\"u}cken}
	\state{Germany}
	\postcode{66123}
}
\email{rzayer@mpi-inf.mpg.de}

\begin{abstract}

In recent years, mesh subdivision---the process of forging smooth free-form surfaces from coarse polygonal meshes---has become an indispensable production instrument.
Although subdivision performance is crucial during simulation, animation and rendering, state-of-the-art approaches still rely on serial implementations for complex parts of the subdivision process.
Therefore, they often fail to harness the power of modern parallel devices, like the graphics processing unit (GPU), for large parts of the algorithm and must resort to time-consuming serial preprocessing.
In this paper, we show that a complete parallelization of the subdivision process for modern architectures is possible.
Building on sparse matrix linear algebra, we show how to structure the complete subdivision process into a sequence of algebra operations.
By restructuring and grouping these operations, we adapt the process for different use cases, such as regular subdivision of dynamic meshes, uniform subdivision for immutable topology, and feature-adaptive subdivision for efficient rendering of animated models.
As the same machinery is used for all use cases, identical subdivision results are achieved in all parts of the production pipeline.
As a second contribution, we show how these linear algebra formulations can effectively be translated into efficient GPU kernels. 
Applying our strategies to  $\sqrt{3}$, Loop and Catmull-Clark subdivision shows significant speedups of our approach compared to state-of-the-art solutions, while we completely avoid serial preprocessing.
\end{abstract}


\maketitle

\input{introduction}
\input{related_work}
\input{method}

\input{optimization}
\input{results}

\input{conclusion}

\bibliographystyle{ACM-Reference-Format}
\bibliography{bibliography}

\input{appendix}

\end{document}

%% file: introduction.tex
\section{Introduction}
\label{sec:introduction}

\begin{figure}[t]
	\includegraphics[width=0.7\linewidth]{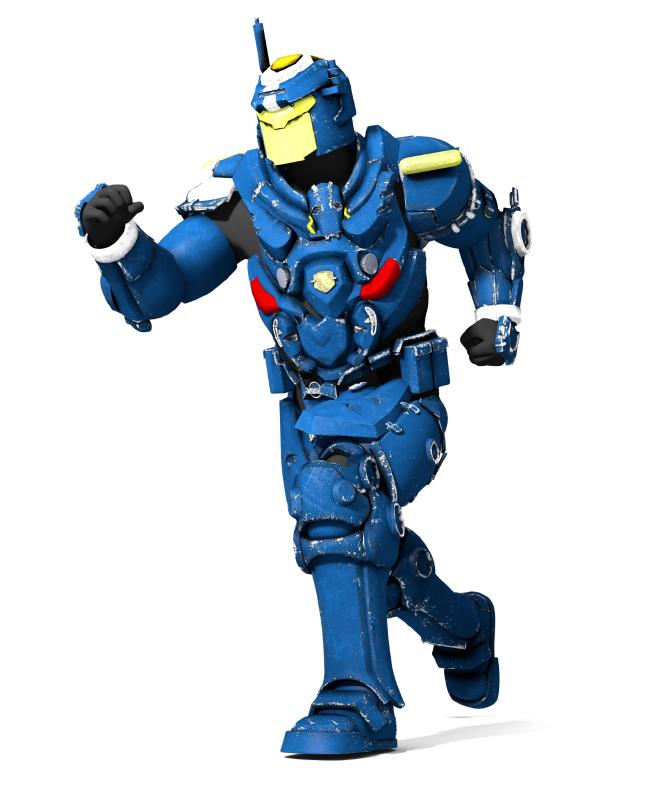}
	\caption{The control mesh of the ArmorGuy (courtesy of DigitalFish) subdivision model consists of 9k faces and 10k vertices and features a considerable number of creases. Using our approach, the refined mesh at level six (35M faces, 35M vertices) can be computed in ~40ms without any preprocessing.}
	\label{fig:teaser}
\end{figure}

Mesh subdivision is a ubiquitous method to generate free-form surfaces from a coarse control mesh, as shown in Figure~\ref{fig:teaser}.
Subdivision surfaces have now been a research topic for over four decades. 
However, their efficient evaluation still poses a challenge on modern parallel architectures.
During computation, the control mesh undergoes a series of averaging, splitting and relaxation operations, which complicates efficient parallel implementation and data management.

In contrast, serial subdivision implementations traditionally rely on mesh representations based on linked lists, \eg, winged-edge representations~\cite{Baumgart1972}.
Changes to the topology in such a data structure requires careful pointer updates to preserve consistency.
Computations in the local neighborhood of mesh vertices---which are essential in subdivision---require pointer chasing.
While those operations are efficient on the CPU, modern parallel devices, like the graphics processing unit (GPU), are faced with unbalanced workloads, synchronization issues, and scattered memory accesses---all of which significantly hurt performance on the GPU.

To bring subdivision to parallel devices, approaches usually  either split the mesh into patches that can be subdivided independently~\cite{Bolz2002, Bolz2003,Shiue2005,Patney2009} or  carry out the bulk of the subdivision process on the CPU and only perform simple operations on the GPU~\cite{Niessner2012,OpenSubdiv}.
While splitting a mesh seems appealing for parallelization at first, the approach entails a series of issues.
First, the per-patch workload and operations depend on the local topology, which leads to execution divergence between the executing entities.
Second, the border between patches needs to be duplicated.
Third, cracks might be introduced between patch boundaries due to floating point problems.
And fourth, re-patching and workload distribution might be required as the model gets subdivided recursively.

The alternative---using the CPU to precompute subdivision tables and only mixing coarse mesh vertices on the GPU---seems an ideal solution for parallel rendering of animated meshes.
However, they do not solve the challenge of parallelization of the subdivision process, but rather build on the fact that serial preprocessing---which may take three orders of magnitude longer than the evaluation---can take place as long as the model has immutable topology.
However, if modeling operations are applied to the mesh or new assets are loaded, preprocessing needs to be applied anew.

Due to the inability of performing the complete subdivision process efficiently in parallel, different approaches are used for various use cases.
When uniform subdivision, \eg, for physics simulation, is required, patch-based parallelization can be used.
During topology-changing modeling operations, only previews of the full subdivision are shown to provide high performance.
After modeling is completed, subdivision tables are used for animation.
Finally, during rendering, partial subdivision or patch-based approaches are used to reduce the workload.
As different approaches also lead to slightly different results, the meshes used for simulation, preview, animation, and rendering may vary at details---a fact bothering many artists.

With Algebra Subdivision, short \emph{AlSub}, we provide a fully parallel and modular subdivision approach, ensuring not only consistent results throughout all application scenarios, but also show significant performance improvements for all of them.
AlSub recasts mesh subdivision into linear algebra operations and is \emph{the first fully GPU-enabled, universally applicable subdivision implementation}. 
We make the following contributions:
\begin{itemize}
	\item We show that with few linear algebra operations optimized for mesh-processing, the entire subdivision process can be described in a compact, self-contained manner suitable for execution on \emph{massively parallel devices} like the GPU.
	\item We show that our sparse linear algebra formalization is \emph{sufficiently general} to describe many existing subdivision schemes such as $\sqrt{3}$, Loop and Catmull-Clark.
	\item We show that the proposed approach \emph{can be easily extended} to support additions to the standard subdivision algorithms, such as sharp and semi sharp creases, displacement mapping and subdivision of selected regions, \eg, for feature adaptiveness or path tracing.
	\item We show that our approach is \emph{modular} such that topological operations can be separated from evaluation, leading to an efficient parallel preprocessing for immuitable topology followed by single matrix-vector product for position updates, \eg, for animation. 
	\item We show that the involved linear algebra operations can be \emph{specialized and optimized} for the use case of mesh processing, leading to highly optimized subdivision kernels.
\end{itemize}

After a brief summary of related work (Section \ref{sec:related_work}), we present the mathematical background and details of our approach for Catmull-Clark subdivision (Section \ref{sec:catclark}). 
The important steps for Loop and $\sqrt{3}$-subdivision are found in the Appendix.
We then highlight how linear algebra operations are translated and optimized for efficient mesh processing kernels (Section \ref{sec:optimization}).
In Section \ref{sec:results}, we show that our implementation outperforms other publicly available production and research implementations such as OpenSubdiv~\cite{OpenSubdiv} and the feature adaptive version of \citeauthor{Niessner2012} \shortcite{Niessner2012}, as well as the patch-based GPU subdivision by \citeauthor{Patney2009}~\shortcite{Patney2009} in their respective domains.
We are currently integrating our approach into the open source modeling and rendering tool Blender~\shortcite{Blender}.

%% file: related_work.tex
\section{Related Work}
\label{sec:related_work}

Subdivision bears some similarity to early ideas in surface fitting in finite element analysis~\cite{Clough1965} and numerical approximation~\cite{Powell:1977:PQA} and it has been honed for geometric modeling through the concerted effort of several pioneering researchers, \eg, \citeauthor{Chaikin:1974:AHS} \shortcite{Chaikin:1974:AHS}, \citeauthor{Doo:1978:SAS} \shortcite{Doo:1978:SAS}, \citeauthor{Doo:1978:BRD} \shortcite{Doo:1978:BRD},
and \citeauthor{Catmull1978} \shortcite{Catmull1978}. 

Subdivision meshes are commonly used across various fields ranging from character animation in feature film production \cite{DeRose1998} to primitive creation for REYES-style rendering~\cite{Zhou2009}, and real-time rendering~\cite{Tzeng2010}.

Mesh subdivision is a refinement procedure which requires data structures capable of providing and updating connectivity information. Commonly used data structures are often variants of the \emph{winged-edge} mesh representations~\cite{Baumgart1972}, like quad-edge \cite{Guibas1985} or half-edge~\cite{Lienhardt1994,Campagna1998}.
While they are well suited for use in the serial setting, parallel implementations suffer from scattered memory accesses, which are particularly harmful to performance. Besides, their storage cost is a limiting factor on graphics hardware.
Compressed alternative formats which were designed for GPU-rendering, like triangle stripes~\cite{Deering1995,Hoppe1999}, do not offer complete connectivity information and are thus not suitable for subdivision.
Patch-based GPU subdivision approaches have thus tried to find efficient patch data structures for subdivision~\cite{Shiue2005,Patney2009}. 

Most recently, a compact sparse matrix mesh representation has been proposed~\cite{Zayer2017}, where mesh processing operations can be expressed as sparse linear algebra and parallelized using linear algebra kernels.
While the principal applicability of parallel matrix operations to mesh processing tasks has been reported, more complex computations have not been attempted. In the same spirit, the effort undertaken by~\citeauthor{Mueller-Roemer2017} \shortcite{Mueller-Roemer2017} for volumetric subdivision attempts to use boundary operators for boosting performance on the GPU. While these differential forms have been used earlier~\cite{Castillo:2005}, their storage cost and redundancies continue to limit their practical scope, especially, as data-sets with millions of elements are now mainstream.

Given the pressing need for high performance subdivision implementations, various vectorization approaches have been proposed.
\citeauthor{Shiue2005} \shortcite{Shiue2005} divide the mesh into fragments which can be subdivided independently on the GPU, which reduces inter-thread communication but introduces redundant data and computations.
Moreover, an initial subdivision step has to be done on the CPU.
Subdivision tables have been introduced to efficiently reevaluate the refined mesh after moving  control mesh vertices~\cite{Bolz2002}.
However, the creation of such tables requires a symbolic subdivision, whose cost is similar to a full subdivision.
Similarly, the pre-computed eigenstructure of the subdivision matrix can be used for direct evaluation of Catmull-Clark  surfaces~\cite{Stam1998}.

To avoid the cost induced by exact subdivision approaches, approximation schemes have been introduced.
\citeauthor{Peters2000} \shortcite{Peters2000} proposed an algorithm that transforms the quadrilaterals of a mesh into bicubic Nurbs patches.
While the resulting surface is tangent continuous everywhere, the algorithm imposes restricting requirements on the mesh.
The approach of \citeauthor{Loop2008} \shortcite{Loop2008} approximates the Catmull-Clark subdivision surface in regular regions using bicubic patches.
Irregular faces still require additional computations.
Approximations like the aforementioned are fast to evaluate, but along the way, desirable subdivision properties get lost and visual quality deteriorates.
While regular faces can be rendered efficiently by exploiting the bicubic representation using hardware tessellation, irregular regions require recursive subdivision to reduce visual errors~\cite{Niessner2012}.
\citeauthor{Schafer2015} \shortcite{Schafer2015} took the idea one step further and enabled different subdivision depths for irregular vertices in a mesh. 
\citeauthor{Brainerd2016} \shortcite{Brainerd2016} improved upon these results by introducing subdivision plans.
Beyond classical subdivision, several extensions have been proposed to allow for  meshes with boundary \cite{Nasri1987}, sharp creases \cite{DeRose1998}, feature based adaptivity~\cite{Niessner2012}, or displacement mapping \cite{Cook1984,Niessner2013}.

Our approach avoids the aforementioned shortcomings and requires neither CPU preprocessing nor expensive mesh data structures.
At the top level, it can be formalized mathematically in the concise language of linear algebra and hence the ensuing algorithms are easy to understand and modify without knowledge of the underlying numerical kernels. 
This level also reveals the modular nature of our approach through which it can be adapted for various use cases.
At the lower level, our formalization discloses numerical patterns across subdivision steps through which we streamline the associated kernels and increase performance. 

%% file: method.tex
\section{Sparse Linear Algebra Subdivision}
\label{sec:catclark}
Given the generality and popularity of the Catmull-Clark subdivision scheme, we  will use it to walk through the algorithmic development of our method. In Appendix \ref{sec:sqrt3} and \ref{sec:loop}, we briefly show how the same ideas apply to Loop and $\sqrt3$ subdivision.

\subsection{Classical Formulation}

The Catmull-Clark Subdivision scheme offers a generalization of bicubic patches to the irregular mesh setting~\cite{Catmull1978}. It can be applied to polygonal faces of arbitrary order and always  produces quadrilaterals regardless of the input.
Figure \ref{fig:CC_scheme} outlines the four steps of a Catmull-Clark subdivision iteration:

\begin{figure}[ht]
	\includegraphics[clip, trim =0cm 6cm 0cm 6cm, width=0.48\textwidth]{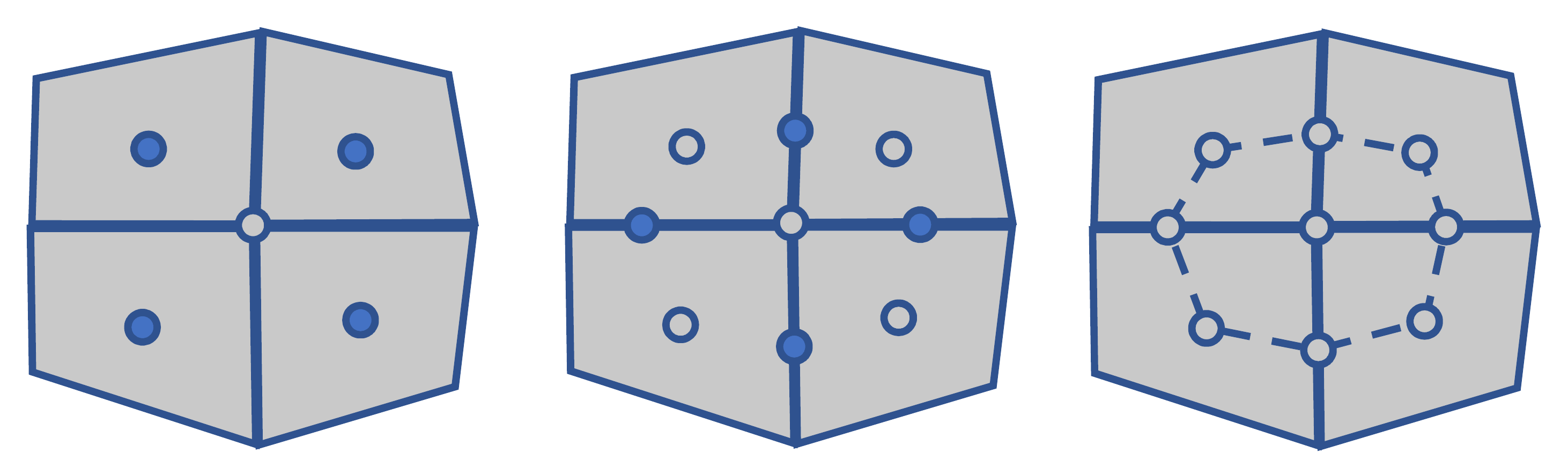}
	\caption{The Catmull-Clark scheme inserts face-points (left), edge-points (center), and creates new faces by connecting face-points, edge-points and the original central point whose location is updated in a smoothing step (right).}
	\label{fig:CC_scheme}
\end{figure}

\paragraph{1. Face-point calculation:} For an arbitrary polygonal face $i$ of order $c_i$, the position of face-point $f_i$ is set to the barycenter of the polygon

\begin{equation}
	f_i = \frac{1}{c_i}\sum_{j=1}^{c_i}p_j
\end{equation}
where $p_j$ are the face vertices.

\paragraph{2. Edge-point calculation:} For each edge $p_kp_l$, a new edge-point is introduced as the average of the endpoints $p_k$ and $p_l$ and the face-points $f_r$ and $f_s$ corresponding to the two faces bordering the edge:

\begin{equation}
	e_{k,l} = \frac{1}{4}\left(p_k + p_l + f_r + f_s\right).
\end{equation}

\paragraph{3. Vertex update:} To produce smooth results, the original vertex location has to be updated using a linear combination of its old position, the edge-mid-points of all incident edges and the surrounding face-points

\begin{equation}
	S(p_i) = \frac{1}{n_i}\left(\left(n_i - 3\right) p_i+\frac{1}{n_i}\sum_{j = 1}^{n_i}f_j + \frac{2}{n_i} \sum_{j = 1}^{n_i}\frac{1}{2}\left(p_i + p_j \right) \right)\text{,}
	\label{eq:pos_update}
\end{equation}

where $n_i$ is the vertex's valence, $f_j$ are the face-points on adjacent faces and $p_j$  the vertices in the $1$-ring neighborhood of $p_i$.

\paragraph{4. Topology refinement:} 
New edges are inserted that connect the face-point to the face's edge-points, splitting each face of order $c$ into as many quadrilaterals.

\paragraph{Boundaries:} Catmull-Clark subdivision can also be used on meshes with boundary.
Edge-points on boundary edges are placed on the edges' mid-points.
Boundary vertex positions $p_i$ are only influenced by adjacent boundary vertices
\begin{equation}
	S(p_i) = \frac{3}{4}p_i + \frac{1}{8}(p_{i-1} + p_{i+1})\text{.}
	\label{eq::CC_boundary}
\end{equation}

\subsection{Linear Algebra Formulation}
\label{sec:cc_linalg}
To derive our linear algebra formulations we use the sparse mesh matrix $\mathcal{M}$~\cite{Zayer2017} as mesh representation.
Each column in $\mathcal{M}$ corresponds to a face.
Row indices of non-zero entries in a column correspond to the face's vertices and the values reflect the cyclic order of the vertex in the face.
Throughout this exposition, we will extend and make use of the action map notation~\cite{Zayer2017}.
The action map notation is used to express alterations of the classical behavior of sparse matrix-vector multiplication (SpMV) and sparse matrix-matrix multiplication (SpGEMM), replacing the core multiplication with alternative operations.
In this way, compact formulations of a sequence of operations are possible.
Additionally, they also hint at efficient implementations as the matrix algebra captures data movement, while action maps capture the actual operations to be carried out. 
We will revisit these facts when discussing an efficient implementation.
We explain the action map notations where they first appear in the following derivation.

\paragraph{Face-point calculation:} 
To compute the barycenters, face orders can be obtained using an action mapped SpMV
\begin{equation}
 \mathbf{c} = \underset{ val\rightarrow 1}{\mathcal{M}^T\mathbf{1}}\text{;}
\label{eq:fo}
\end{equation}
where $\mathbf 1$ is a vector of ones spanning the range of the faces. The mapping below the multiplication indicates that the non-zero values of $\mathcal{M}^T$ will be replaced by a $1$ during multiplication. This simply yields the number of vertices of each face.

The face-points can then be obtained using the mapped SpMV
\begin{equation}
	\mathbf{f} = \underset{ val_{i,\ast}\rightarrow \frac{1}{c_i}}{\mathcal{M}^T \mathbf{P}}\text{,}
	\label{eq:spla_fp}
\end{equation}
where $\mathbf P$ is the array of all vertex coordinates.
In this case, the entries read from the matrix are used as indices into a one dimensional map.
Every non-zero value $val_{i,\ast}$ in $\mathcal{M}^T(i,\ast)$ is mapped to the reciprocal of the order of face $i$.

For a quadrilateral mesh, the SpMV simplifies to
\begin{equation}
\mathbf{f} = \underset{val \rightarrow \frac{1}{4}}{\mathcal{M}^T\mathbf{P}}\text{.}
\end{equation}

\paragraph{Edge-point calculation:}

The computation of edge points requires assigning unique indices to mesh edges.
Such an enumeration can be obtained from the upper (or lower) triangular part of the adjacency matrix associated with the undirected graph of the mesh.
With standard sparse matrix machinery this matrix can be created, for instance, by first computing the adjacency matrix of the oriented mesh graph and then summing it with its transpose, to account for meshes with boundaries.
In view of our high performance goals, this is not a viable approach since it requires additional data creation (transpose), and more importantly, matrix assembly which is notoriously challenging on parallel platforms.

With action maps this can be conveniently encoded as
\begin{equation}
	E = \underset{\lbrace Q_{c} + Q_{c}^{c-1}\rbrace  \lbrack \lambda \rbrack}{\mathcal{M}\mathcal{M}^T}.
	\label{eq:CC_E}
\end{equation}

For the computation of $E$, the two circulant matrices $Q_{c}$ and its power $Q_{c}^{c-1}$, where $c$ is the face order, are combined to capture the counterclockwise and clockwise orientation inside a given face.
In this context, action maps in SpGEMM are small matrices.
Whenever a collision between entries of two matrices occurs during the SpGEMM, the non-zero values are used as indices into the map.
The map value is then used as a result of the collision (instead of the product of the two values).
Therefore, entries in the result matrix of the mapped SpGEMM are a sum of map values.

For quads, $Q_4$ captures the CCW and $Q_4^3$ the CW adjacency.
\begin{equation}
Q_4 =
\kbordermatrix{
	& 1 & 2 & 3 & 4\\
	1 & 0 & 1 & 0 & 0\\
	2 & 0 & 0 & 1 & 0\\
	3 & 0 & 0 & 0 & 1\\
	4 & 1 & 0 & 0 & 0
}\text{,\quad}
Q_4^3 =
\kbordermatrix{
	& 1 & 2 & 3 & 4\\
	1 & 0 & 0 & 0 & 1\\
	2 & 1 & 0 & 0 & 0\\
	3 & 0 & 1 & 0 & 0\\
	4 & 0 & 0 & 1 & 0
}\text{;}
\end{equation}
These maps do not have to be created explicitly, as their entries can be computed on demand.
This is particularly useful, when the face types vary within a mesh:
\begin{equation}
Q_c^r\left(i,j\right) =
\begin{cases}	
1 & if \quad j = \left(\left(i+r-1\right)\mod c\right) + 1 \\	
0 & else	
\end{cases}
\end{equation}

We extended the original action map notation by functions, $\lambda$ in Equation \ref{eq:CC_E}, which are called each time a collision between elements $\mathcal{M}(i,k)$ and $\mathcal{M}^T(k,j)$ happens.
It performs the map lookup and, depending on the map value, computes the result of a collision:

\begin{equation}
	\lambda(i,j) = Q(i,j)
	\label{eq:uniqEdgeCC}
\end{equation}
If the map entry is non-zero, the vertices $p_i$ and $p_j$ are connected to each other within a face $k$.
Unique indices for edges can easily be generated by enumerating the non-zeros in the upper triangular part of the matrix $E$.

To complete the computation of edge-points, faces adjacent to a given edge are required.
For this purpose, a secondary matrix $F$ can be used. This matrix has the same sparsity pattern as the adjacency matrix of the oriented graph of the mesh but each non-zero entry $i,j$ stores the index of the face containing the edge $p_ip_j$.
It can be similarly constructed by matrix multiplication such that whenever the action map returns a non-zero for a collision between elements $\mathcal{M}(i,k)$ and $\mathcal{M}^T(k,j)$, the face index $k$ is stored in $F(i,j)$.
\begin{equation}
F = \underset{\lbrace Q_{c}\rbrace  \lbrack \gamma \rbrack}{\mathcal{M}\mathcal{M}^T}
\label{eq:F}
\end{equation}
with the function
\begin{equation}
\gamma(i,j,k) =
\begin{cases}	
k & if \quad Q_c = 1 \\	
0 & else	
\end{cases}
\end{equation}
Hence, for each edge $p_ip_j$ in the mesh, its unique edge index is known from $E$ and the two adjacent faces are $F(i,j)$ and $F(j,i)$.
The edge-point position can then be computed.

\paragraph{Vertex update:}

The position update in Equation \ref{eq:pos_update} can be conveniently rewritten as
\begin{equation}
	S(p_i) =
	\underbrace{\vphantom{\frac{1}{n_i^2}\sum_{j = 1}^{n_i}p_j}\left(1 - \frac{2}{n_i}\right) p_i}_{s_1} +
	\underbrace{\frac{1}{n_i^2}\sum_{j = 1}^{n_i}p_j}_{s_2} +
	\underbrace{\frac{1}{n_i^2}\sum_{j = 1}^{n_i}f_j}_{s_3} \text{,}
\end{equation}
such that the update can be split into three summands. Vertex valencies can be obtained globally as the vector
\begin{equation}
\mathbf{n} = \underset{ val\rightarrow 1}{\mathcal{M}\mathbf{1}}\text{.}
\label{eq:vo}
\end{equation}
$s_1$ involves only the original position and can be calculated in the customary ways.

The second summand, $s_2$, sums the 1-ring neighborhood of the vertex.
This is done using the matrix $F$, which has the same sparsity pattern as the vertex-vertex adjacency matrix without diagonal, in the mapped SpMV
\begin{equation}
	s_2 = \underset{ val_{i,\ast} \rightarrow \frac{1}{n_i^2}}{F\mathbf{P}} \text{.}
\end{equation}
The last term sums the face-points on faces adjacent to the vertex and is computed via
\begin{equation}
	s_3 = \underset{ val_{i,\ast} \rightarrow \frac{1}{n_i^2}}{\mathcal{M}\mathbf{f}} \text{.}
\end{equation}

\paragraph{Topology refinement:}
A new face consists of one (updated) vertex of the parent, its face-point and two edge-points.
To capture a face in a mesh matrix $\mathcal{M}$, a column representing the polygon has to be added to the matrix.
The non-zero locations in each new column correspond to the referenced vertices.
Thus, for creating the topology of the subdivided mesh, the computed vertices must be referenced accordingly and $\mathcal{M}$ of the refined mesh matrix has to be assembled.

A column $\mathcal{M}(\ast,r)$ in the control mesh matrix is replaced by $c_r$ columns in the refined mesh matrix where $c_r$ is the order of the face.
The indices of the original vertices are already known; the index of the face-point on $f_r$ is $\lvert v \rvert + r$, where $\lvert v \rvert$ is the number of mesh vertices.
The indices of the two edge-points can be determined by fetching the edges' indices from $E$ and incrementing them by $\lvert v \rvert + \lvert f \rvert$, where $\lvert f \rvert$ is the number of mesh faces.
Performing these steps for all new faces yields $\mathcal{M}$ for the refined mesh.

\begin{figure}
	\centering
	\includegraphics[width=\linewidth]{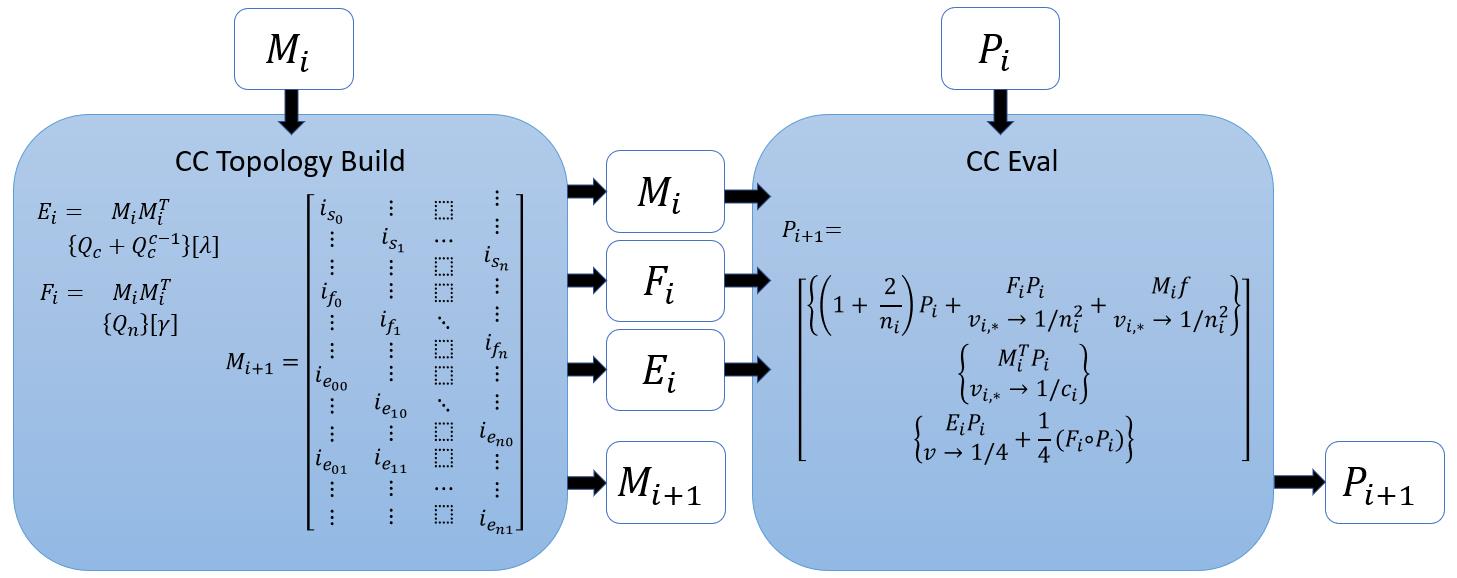}
	\caption{The core steps of one iteration of our Catmull-Clark subdivision are split into a build step, which is concerned with the topological operations, and an evaluation step, which receives information from the build step and the base mesh vertices.}
	\label{fig:module_core}
\end{figure}

The combination of all steps mentioned above is outlined in Figure~\ref{fig:module_core}.
As can be seen, one iteration can be split into a build and an eval step.
The build step takes the current mesh matrix and generates $F$, $E$ and mesh matrix of the subdivided mesh.
In this way, all topology related operations are carried out by the build step.
The eval step, receives the matrices $F$ and $E$ as well as the mesh matrix and vertex positions from the last iteration.
It then carries out the mapped SpMVs to generate the new vertex locations.
 
\begin{figure}
	\centering
	\includegraphics[width=0.8\linewidth]{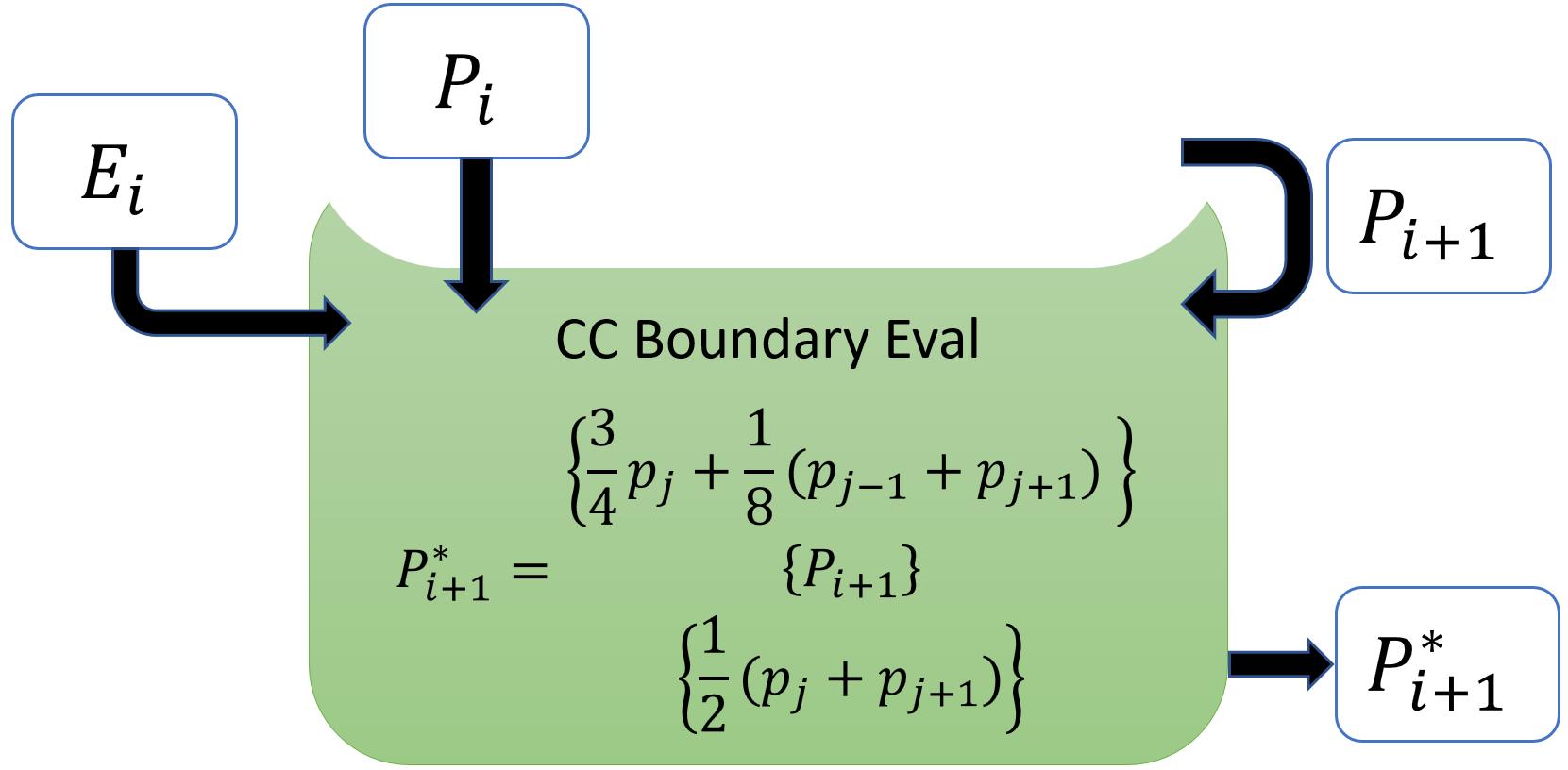}
	\caption{Boundary evaluation is captured by our approach as a simple additional step after evaluation, and can be seen as a module appended to the standard evaluation.}
	\label{fig:module_boundary}
\end{figure}

\paragraph{Boundaries:} In practice, meshes often feature boundaries, which need to be treated using specialized subdivision rules.
AlSub handles boundary meshes in a build and repair fashion.
First, the refined vertex data is computed as usual.
In a subsequent step, boundary vertices can be conveniently identified from $E$ as entries which have a value of $1$, and are repaired in parallel according to Equation \ref{eq::CC_boundary}.
Edge-points on edges connecting external vertices are set to the edge-mid points.
Their indices can again be obtained from the enumeration of the non-zeros in the upper triangular part of $E$.
Adding boundaries to our approach essentially forms an additional step after the default evaluation, as shown in Figure~\ref{fig:module_boundary}.

\label{sec:cc_creases}
\begin{figure}
	\centering
	\includegraphics[width=.98\linewidth]{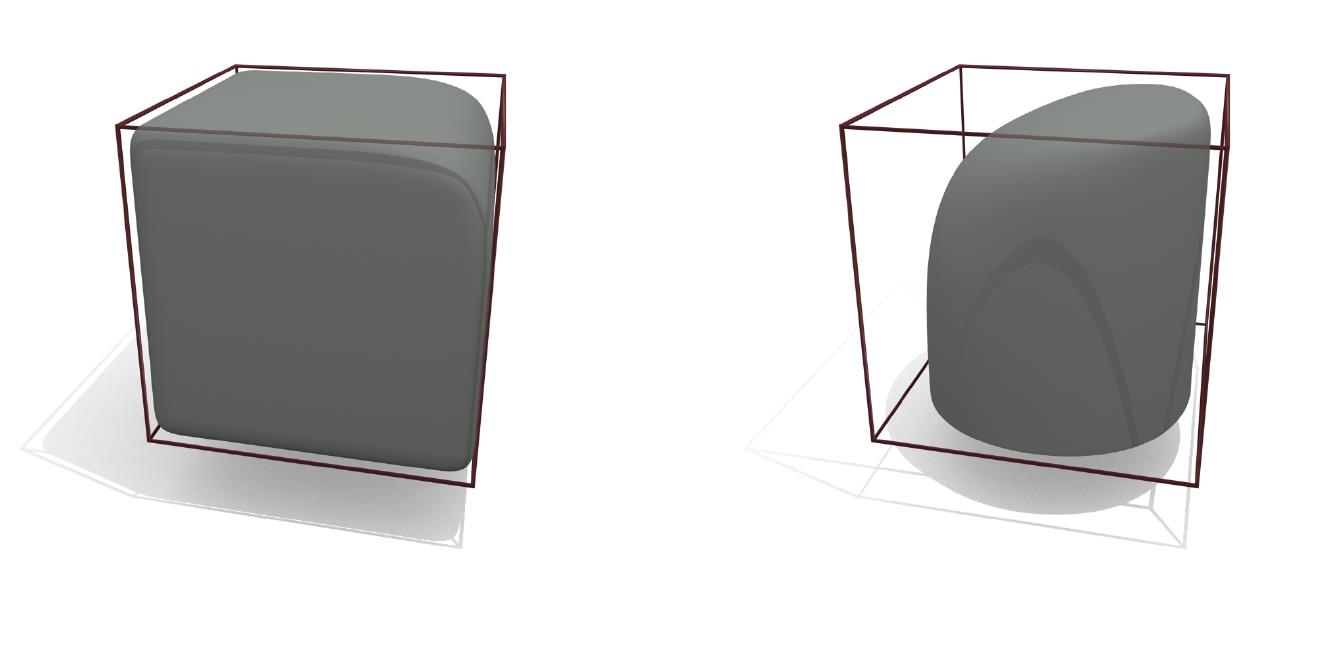}
	\caption{
		Our approach naturally supports extensions, such as semi-sharp and infinitely sharp creases, as shown here on a cube.
	}
	\label{fig:creases}
\end{figure}

\subsection{Creases}
Sharp and semi-sharp creases have become indispensable in subdivision surface modeling to describe piecewise smooth and tightly curved surfaces~\cite{DeRose1998}, \cf Figure~\ref{fig:creases}.
Creases are edges that are tagged by a (not necessarily) integer sharpness value and updated according to a special set of rules during subdivision.
As the general computation of creases is beyond the scope of this paper we refer the reader to \citeauthor{DeRose1998}~\shortcite{DeRose1998} for a detailed description, while we only present their treatment as sparse matrix linear algebra.
 
To support creases, we use a sparse symmetric crease matrix $C$ of size $\lvert v \rvert \times \lvert v \rvert$.
The entry $C(i,j) = \sigma_{ij}$ holds the sharpness value of the crease between vertices $i$ and $j$.
To calculate the position of crease vertices and edge points, the crease valency $\mathbf{k}$, \ie, number of creases incident to a crease vertex
\begin{equation}
\mathbf{k} = \underset{val \rightarrow 1}{C\mathbf{1}}
\label{eq:CC_crease_valency}
\end{equation}
and the vertex sharpness $\mathbf{s}$, \ie, average over all incident crease sharpnesses
\begin{equation}
\mathbf{s} = \underset{val_{i,j} \rightarrow \frac{val_{i,j}}{k_i}}{C\mathbf{1}}
\label{eq:CC_crease_vsharpness}
\end{equation}
need to be determined, which we complete using the same SpMV with two different maps.
With the computed vectors $\mathbf{k}$ and $\mathbf{s}$ and the already available adjacency information in $E$, we correct crease vertices in parallel using the rules provided by \citeauthor{DeRose1998}~\shortcite{DeRose1998}.
After each iteration of subdivision a new crease matrix is created, that holds the updated sharpness values for the subdivided creases.
This crease inheritance is performed in two steps:
(1) The sparsity pattern is determined by updating crease values according to a variation of Chaikin's edge subdivision algorithm~\cite{Chaikin:1974:AHS} that accounts for decreasing sharpness values~\cite{DeRose1998}

\begin{equation}
	\sigma_{ij} = \max \lbrace \frac{1}{4} \left( \sigma_i + 3 \sigma_j \right) - 1, 0 \rbrace
\end{equation}
\begin{equation}
	\sigma_{jk} = \max \lbrace \frac{1}{4} \left( 3 \sigma_j + \sigma_k \right) - 1 , 0 \rbrace
\end{equation}

where $\sigma_i$, $\sigma_j$ and $\sigma_k$ are sharpness values of three adjacent parent creases $i$, $j$ and $k$. 
$\sigma_{ij}$ and $\sigma_{jk}$ are the sharpness values of the two child creases of $j$.
To allocate the memory for the new crease matrix, the number of resulting non-zero sharpnesses in each of it's columns is counted.
(2) The inherited crease matrix is subsequently filled with the remaining non-zero sharpness values.
A similar routine as in the first step is used which now fills in the updated non-zero crease values and their indices.
If all crease sharpnesses decreased to zero, the subsequent subdivision steps are carried out as for a smooth mesh.

\begin{figure}
	\centering
	\includegraphics[width=\linewidth]{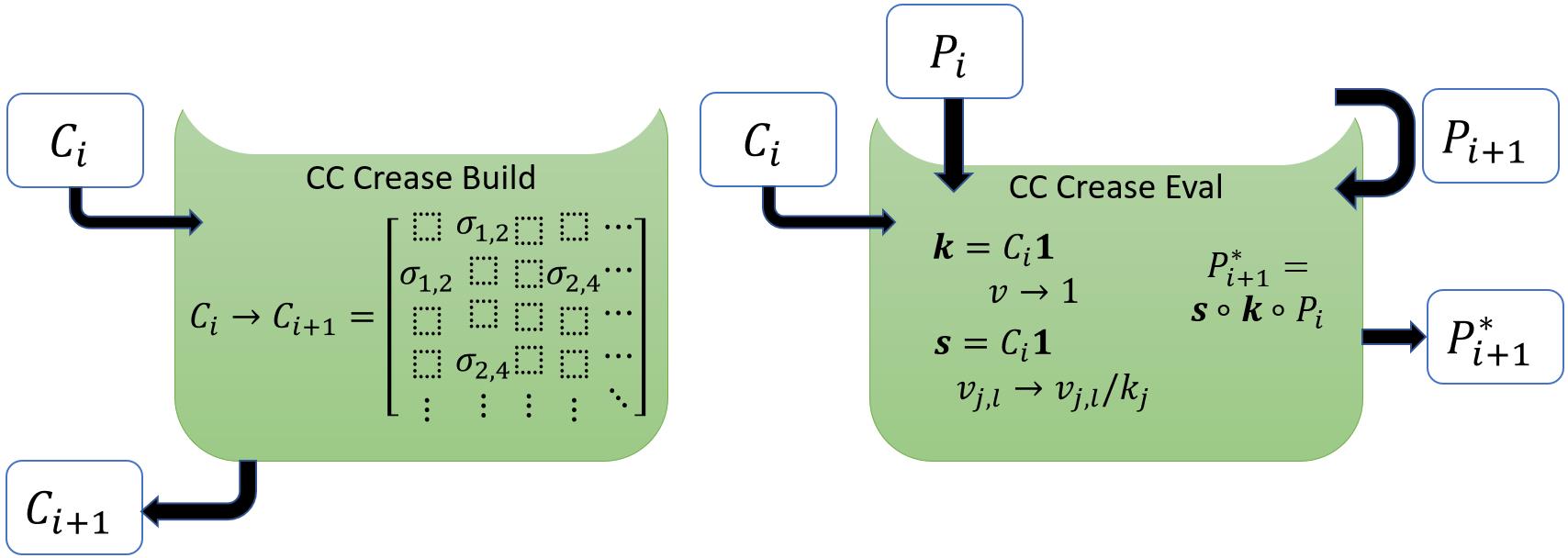}
	\caption{Creases are modeled as a single sparse crease matrix in our approach, which is updated each iteration. During evaluation creases simply overwrite the vertex positions from the previous subdivision step. }
	\label{fig:module_creases}
\end{figure}

The addition of creases to the evaluation process is outlined in  Figure~\ref{fig:module_creases}.
The core of the subdivision process simply remains the same; the crease matrix is created additionally in every subdivision iteration.
During evaluation, vertices influenced by a crease are reevaluated and overwrite the output vertices.

\subsection{Selective and Feature Adaptive Subdivision}
\label{sec:cc_feature_adapt}
The machinery of linear algebra cannot only be used to describe uniform subdivision, but also selective processing, which is interesting for hardware supported rendering~\cite{OpenSubdiv, Niessner2012} and spatially coherent path- and ray-tracing.
As example, consider feature adaptive subdivision.
In a quadrilateral mesh, faces with a consistent vertex valence of four can also be represented as bicubic patches and their limit surface can be evaluated directly.
Thus, for efficiency reasons, one may want to only recursively subdivide around vertices with a valence different from four---around extraordinary vertices.

Using our scheme, extraordinary vertices are easily identified from Equation \ref{eq:vo}, \ie, where the valency is $\neq 4$.
To identify the regions around the extraordinary vertices, we start with a vector $\mathbf{x_0}$ spanning the number of vertices.
$\mathbf{x_0}$ is $0$ everywhere except for extraordinary vertices, where it is $1$.
To determine the surrounding faces, we propagate this information with the mesh matrix $\mathcal{M}$.

First, the neighboring faces are determined as the non-zeros of the vector

\begin{equation}
\mathbf{q_i}= \mathcal{M}^T \mathbf{x_i}
\end{equation}
and their vertices can be revealed as the non-zero entries resulting from the product
\begin{equation}
\mathbf{x_{i+1}} = \mathcal{M}\mathbf{q_i}\text{.}
\end{equation}
This also shows that the adjacency matrix can be obtained from the mapped mesh matrix product and that the power of the adjacency matrix reflects the neighborhood order around a vertex.

Using the information from above, we construct the matrix $X_i$, which has columns equal to the number of vertices in the input mesh and rows equal to the number of vertices that are selected for subdivision.
The entries of $X_i$ correspond to an identity matrix with deleted rows due to $\mathbf{x_{i+1}}$.
The extraction of the vertex data is then performed by the SpMV
\begin{equation}
\mathbf{P^\prime_i} = X_i\mathbf{P_i}\text{.}
\end{equation}

To extract the mesh topology, the matrix $\mathring{X}_i$---analogue to $X_i$---is created from the information acquired in the propagation step.
$\mathring{X}_i$ has rows equal to the number of faces in the original mesh and columns equal to the number of faces in the extracted mesh.
$\mathring{X}_i$ can again be created from the identity matrix by, in contrast to $X_i$, deleting \emph{columns} corresponding to faces that should be disregarded during extraction.
This information is readily available in $\mathbf{q_i}$.
The extracted mesh matrix is then determined via

\begin{equation}
	 \mathcal{M}^\prime = X_i\mathcal{M}\mathring{X}_i
\end{equation}
 
\begin{figure}
	\centering
	\includegraphics[width=0.82\linewidth]{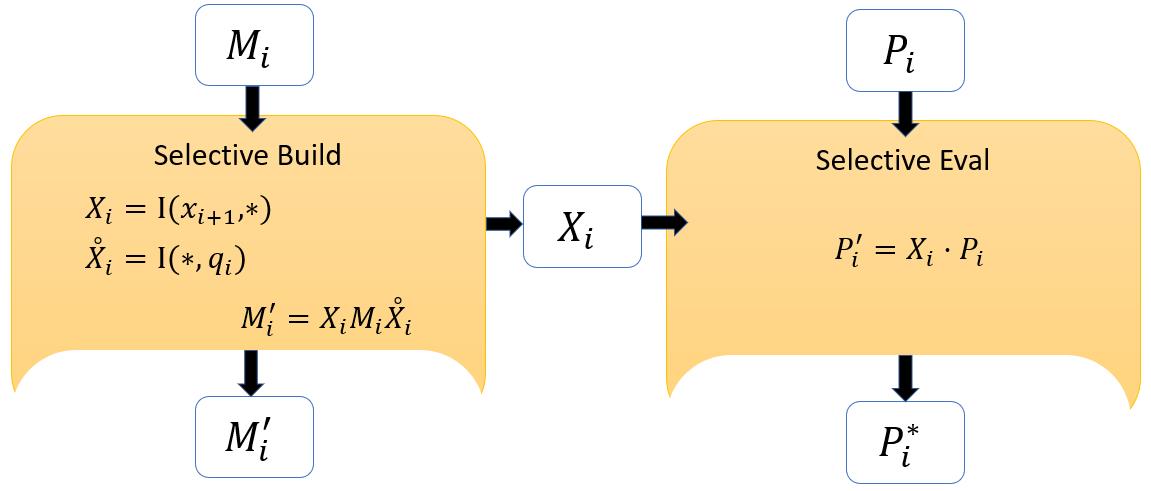}
	\caption{Selective or feature adaptive subdivision is modeled by the extraction matrices $X_i$ and $\mathring{X}_i$, which are generated by identifying the surrounding of selected vertices. These matrices are applied to $\mathcal{M}$ and the vertex data to reduce the subsequent operations to the extracted regions.}
	\label{fig:module_selective}
\end{figure}

Selective subdivision can be seen as a module added before the major subdivision step, as shown in Figure~\ref{fig:module_selective}.

\subsection{Other effects}
\begin{figure}
	\includegraphics[width=0.98\linewidth]{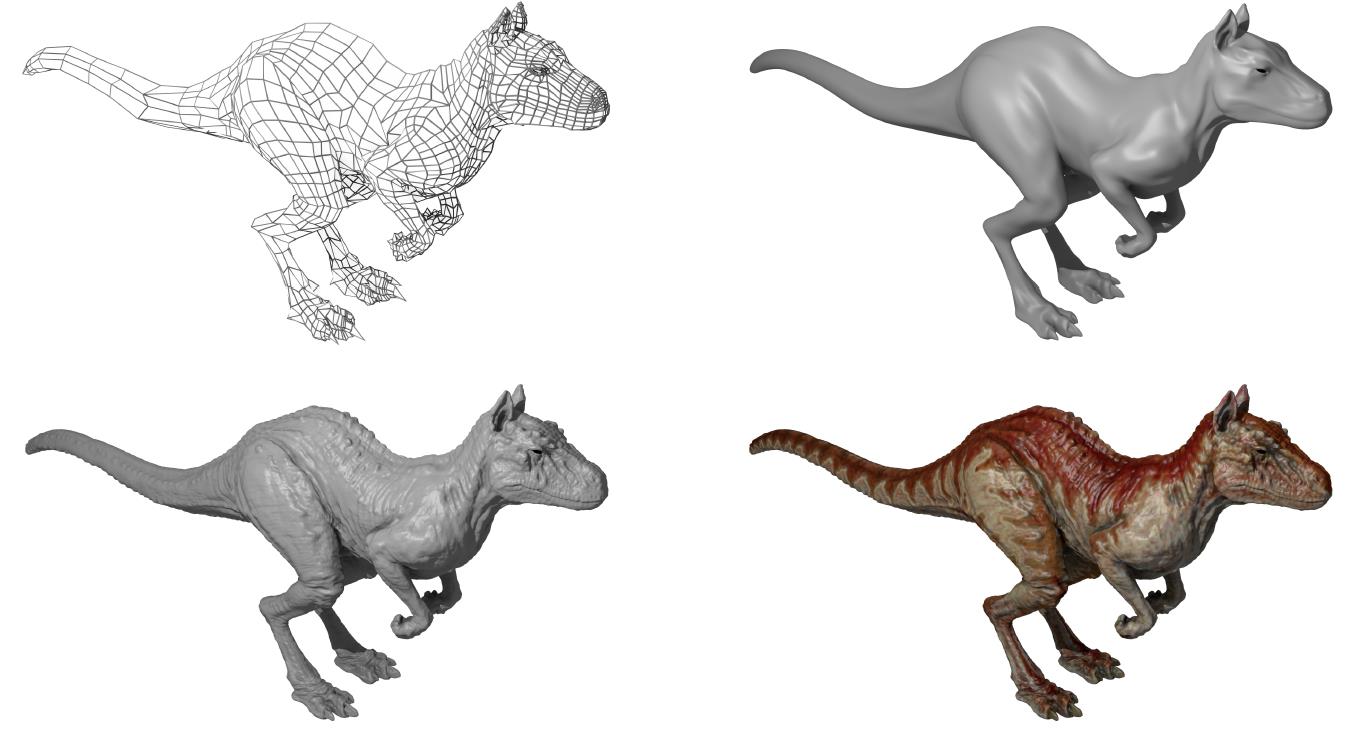}
	\caption{AlSub is capable of subdividing a coarse control mesh (top, left) instantaneously to a dense and smooth refined mesh (top, right). Displacement (bottom, left) and texture (bottom, right) mapping are possible naturally without any additional effort.}
	\label{fig:CC_displ}
\end{figure}

Note that displacement mapping and hierarchical edits \cite{Forsey1988} are also straight forward in our approach, as we have access to the vertex data after each iteration and can arbitrarily modify it.
An example displacement and texture mapped subdivision model can be seen in Figure \ref{fig:CC_displ}.

\subsection{Modes of Operation}
\label{sec:cc_modes}
As already hinted in the previous sections, AlSub is a modular system which allows for dynamic adaption to the requirements of different applications.
However, we distinguish two main categories: dynamic and static topology of the control mesh.

\paragraph*{Dynamic topology} Dynamic topology is ubiquitous in 3D modeling and CAD applications during the content creation process.
Faces, vertices and edges are frequently added, modified and removed which poses a great challenge to many existing approaches that rely on expensive preprocessing, as it has to be repeated on every topological update.
This fact has led to the use of different subdivision approaches for model preview and production rendering causing discrepancies between the two images.
Due to the efficiency of our complete approach, we can avoid any preprocessing and alternate between what we call build steps and eval steps, computing one complete subdivision step before the next.
As additional data like $F_i$ and $E_i$ are only required for one step, memory requirements are relatively low in our approach.

\begin{figure}
	\centering
	\includegraphics[width=\linewidth]{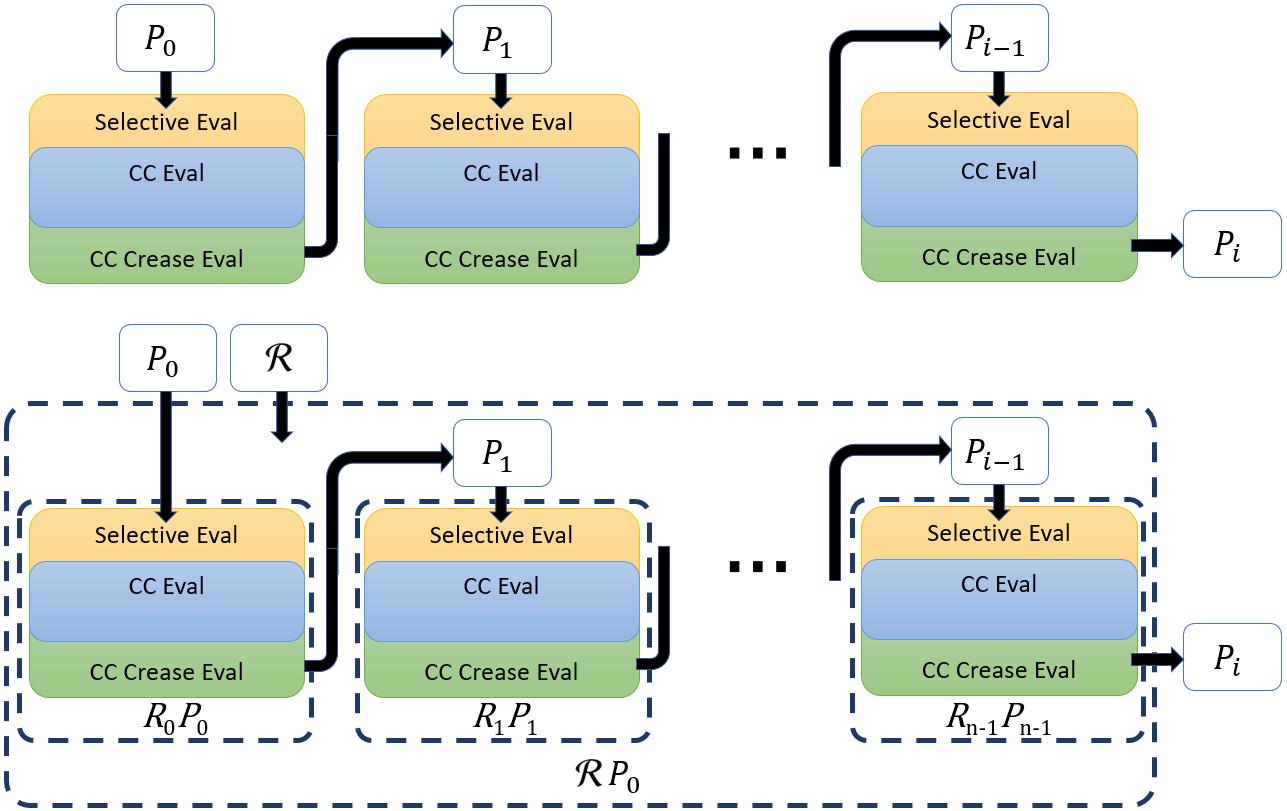}
	\caption{If the topology of the mesh is static only the right-hand side of the modules has to be processed as shown here for adaptive subdivision of a closed mesh (no boundary handling) with creases, using the iterative eval (top) or the single SpMV eval (bottom).}
	\label{fig:CC_singleSpMV}
\end{figure}

\paragraph*{Static topology} Static topology is common, \eg, in production rendering applications, where only vertex attributes, \eg positions, change over time but the mesh connectivity is invariant.
Subdivision algorithms make heavy use of adjacency information.
The fact that this information can be prepared upfront and does not have to be re-computed every frame, reduces the overall production time.
In AlSub, all computations dealing with mesh connectivity are factored into a \emph{build step}, that is executed only once before the mesh is subdivided many times, \ie, generating all $F_i$, $E_i$ and $M_{i+1}$ (as well as $X_i$ and $C_i$ in case of selective subdivision and creases).
Only in the \textit{evaluation step}, we process the right side of all modules, as visualized in the top row of Figure~\ref{fig:CC_singleSpMV}.

\paragraph*{Single SpMV evaluation}
Given that each iteration of the evaluation is a sequence of mapped SpMVs, it is also possible to capture the entire sequence in a single sparse matrix $R_i$:
$R_i$ captures the evaluation of a single subdivision step from level $i$ to level $i+1$ .
Each column in $R_i$ corresponds to a vertex at subdivision level $i$ and each row corresponds to one refined vertex at level $i+1$.
A single iteration of subdivision of vertex data can then be done using the SpMV
\begin{equation}
\mathbf{P_{i+1}} = R_{i}\mathbf{P_{i}}\text{.}
\end{equation}
Building these refinement matrices $R_i$ is simple: instead of calculating the refined vertices directly as would usually be done in the evaluation step, the weights are distributed into the matrix.
We do this in a two stage approach, where we first determine the number of non-zero entries in each row and in the second stage rows are populated with indices and weights.

The entire evaluation from the first level to a specific level $i$ can be written as a sequence of matrix vector products as follows:
\begin{equation}
\mathbf{P_i} = R_{i-1} R_{i-2}\ldots R_1 R_0 \mathbf{P_0} = \mathcal{R} \mathbf{P_0} \text{.}
\label{eq:singleSpMV}
\end{equation}

As all matrices involved in Equation \ref{eq:singleSpMV} are independent of the actual vertex data and therefore only depend on the mesh topology and features such as creases, the subdivision matrix $\mathcal{R}$ can be computed in the build step.
That means the \emph{whole evaluation step}, boils down to a \emph{single SpMV}, regardless of the subdivision depth as shown in Figure~\ref{fig:CC_singleSpMV}, bottom row.
This enables optimization techniques for SpMV kernels to be applied to the evaluation step.

%% file: optimization.tex
\section{Optimization of algebraic Operations}
\label{sec:optimization}
The higher level formalization discussed in Section \ref{sec:catclark} can be easily implemented by minor adjustment to standard sparse matrix algebra kernels.
However, the compact action map notation hints that further specialized and optimized implementations of these operations are possible.
Using the knowledge about the structure of the underlying matrices, we exploit the particular computational patterns of these operations and streamline them through efficient and highly optimized GPU kernels.

\subsection{Reduced Mesh Matrix}
We use the Compressed Sparse Column (CSC) matrix format, which is comprised of three arrays.
The first two hold row indices and values of non-zero entries.
The column pointer contains an index to the start of each column in the first two arrays \cite{Saad:1994:SBT}.
If each column in $\mathcal{M}$ has the same number of non-zero entries, \eg quad or triangle mesh, the column pointer can be omitted.
Reordering the row index-value pairs of $\mathcal{M}$ according to the values in each column also renders the value array unnecessary, because the cyclic order of vertices in a face is then implicitly given by the order of their appearance in the row indices array.
The memory requirement of the reduced mesh matrix is therefore equal to that of a face table of the mesh.

As Loop and $\sqrt{3}$ work on triangle meshes and Catmull-Clark only produces quadrilaterals, we apply this optimization to all our kernels, using general kernels only for the first Catmull-Clark subdivision step. 
In this way, we cut down data creation, memory consumption and memory accesses.

\subsection{Implicit mapped sparse matrix-matrix multiplication}
Mapped multiplications of the form
\begin{equation}
A = \underset{\lbrace Q \rbrace [\alpha]}{\mathcal{M}\mathcal{M}^T}
\label{eq:general_mSpGEMM}
\end{equation}
are extensively used in our mathematical formalization for capturing various connectivity information.
Despite the steady improvement of SpGEMM implementations, their cost is still relatively high.
Therefore, it is worthwhile to avoid explicit multiplication if possible.

A close examination of what happens during multiplications as in Equation $\ref{eq:general_mSpGEMM}$ reveals that the result can be directly created from $\mathcal{M}$.
In the computation of $\mathcal{M}\mathcal{M}^T$, each row $r_i = \mathcal{M}(i, \ast)$ is multiplied with each column $c_j = \mathcal{M}^T(\ast,j)$.
Both vectors, $r_i$ and $c_j$, encode one vertex each and have non-zero entries in the locations corresponding to their adjacent faces.
A collision (and with that an invocation of $\alpha$) between two entries $\mathcal{M}(i,k)$ and $\mathcal{M}^T(k,j)$ happens if both are non-zero, meaning that vertices $i$ and $j$ share a face $k$.
Clearly, vertices not part of the same face will never induce a collision and therefore never produce a non-zero entry in the result.
Conversely, each non-zero entry is produced by two vertices sharing a face.

The above findings allow us to restructure the computation as follows:
We parallelize over the columns of $\mathcal{M}$, \ie, the indices of vertices that form a face.
All mapped multiplications that produce a non-zero result happen within the same column of $\mathcal{M}$.
Simply evaluating $\alpha$ for all pairs of vertices of such a column leads to the desired result.
However, as $Q$ and thus $\alpha$ often contain many zero elements, this approach would still lead to wasted computations.
Thus, instead of using $Q$ to lookup the results for pairs, we use it to guide data loading instead.
For a specific vertex $i$, we find those entries in $Q$ that result in non-zero entries, \ie, the vertex offsets in the face that result in non-zero results, and directly load those values and evaluate $\alpha$.
In this way, we only load and invoke $\alpha$ for those pairs of vertices that actually contribute to the output. 

Before the actual multiplication can be carried out as described above, the number of non-zeros of the result needs to be determined, to allocate the arrays for the output matrix.
To this end, we complete a symbolic pass, similar to general SpGEMM implementations.
In parallel for each entry of $\mathcal{M}$, we determine the number of \emph{local} per-face non-zero $\alpha$ invocations for the vertex by counting the non-zeros in the map row corresponding to the vertex's position. 
We atomically accumulate the \emph{global} number of non-zero for each vertex in an array, which then corresponds to the number of non-zero entries in the vertex's column of the result.
A simple parallel scan (cumulative sum) over that array gives the column pointer and the number of non-zeros of the resulting matrix.
It is worth noting that this step can be skipped if each row of the map has the same number of non-zero entries $z_{r_i} = z_r$.
Then the number of non-zero values of the vertex's column in the result is independent of its position in the cyclic order of adjacent faces.
Thus, the invocations on each vertex can be directly calculated as a multiple of the vertex order $z_r \mathbf{n}$.

\subsection{Specialized SpMVs}
Certain patterns in the mapped SpMVs of the higher level formulations can be transformed to specifically tailored and more efficient GPU kernels.

\paragraph{Direct mapped SpMV}
In the mapped SpMV for CSC matrices, multiple threads collaborate to calculate a single element of the result vector.
Parallelization is done over the elements of the vector.
A thread reads a single entry of the input vector and multiplies it with the mapped non-zero elements of the corresponding column.
These intermediate products are directly accumulated in the result vector using an atomic addition operation to avoid race conditions.
We found that this execution pattern is more efficient than an explicit matrix transpose to avoid atomic operations.

\paragraph{Transpose mapped SpMV}
Due to using CSC matrices, the transpose mapped SpMV can be parallelized easily without an explicit transposition.
We use a single thread per output element, which eliminates the need for atomic operations.
Each thread iterates over the non-zero elements of its column, uses the map to substitute them and multiplies each mapped value with the corresponding vector element.
This means that in contrast to the direct mapped SpMV, each thread reads multiple elements from the input vector.

\paragraph{Specializations}
We distinguish between matrix, vector and map-based optimizations.
Depending on the input parameters to the mapped SpMVs, we apply all matching optimization steps to produce more efficient GPU kernels.

If the \textit{input matrix} is in reduced form, every column has the same number of non-zeros, which renders the column pointer obsolete.
This lets us unroll the loop over each column which eliminates costly conditional jump instructions.
Value arrays can also be omitted, because row indices in each column are sorted to reflect the cyclic order of the face.
In both SpMV versions each thread works on one column of the matrix.
If the row indices are 16 Byte aligned, \eg, mesh matrix of a quad mesh, a single vectorized load is used for four row indices, which increases memory performance.

Many of our operations involve multiplication of the mesh matrix with a predefined \emph{input vector}, \eg, a vector of ones.
In this case the reads of vector elements is obsolete and can be omitted, as specialized kernels for that specific input vector can be generated.
In many cases a vector of positions is used in a mapped SpMV with the mesh matrix.
As every input position consists of multiple components the number of threads can be increased such that the multiplication is carried out on a per-component level.
Without loss of generality, consider the case of averaging the vertex positions for each face, \eg, when calculating face-points in the Catmull-Clark scheme.
Each position consists of four components and each column in $\mathcal{M}$ has four non-zero entries.
In this case an SpMV kernel is launched with 16 threads per face, each responsible for a single component of one vertex position.
Each group of 16 consecutive threads then calculates the mapped multiplication of a single column.
As each output component depends on intermediate products of four vertices, efficient SIMD level communication primitives (shuffle instructions on NVIDIA hardware) are used to combine the results.
The result is then written by a single thread to eliminate the need for atomics.

Furthermore, we exploit properties of the \emph{map} to optimize SpMV kernels.
If the map is a constant function, the value of the map can simply reside in shared or constant memory or even in a register to eliminate frequent map lookups.
In the non-transposed case, maps that are constant per column can be handled similarly.

\paragraph{Single SpMV evaluation}
In the adaptive Catmull-Clark implementation we use a single SpMV to subdivide vertex-data from the control mesh to some predefined level.
As the matrix captures the combination of multiple mapped SpMVs, there is usually no common structure to exploit.
However, as the resulting matrix $\mathcal{R}$ is used for a single SpMV, we store it in CSR format for a more efficient row access.
Furthermore, we pad the row indices and value arrays, such that each row is 16 Byte aligned, to enable vectorized loads independent of the row length.
For the same reason we also pad the vertex-data vector.
For evaluation, we assign eight non-zeros to a single thread which performs the multiplication with eight padded entries in the vertex array, \ie, 32 values.
Then, we use SIMD level communication primitives to merge results that correspond to the same row spread across multiple threads.
Finally, we collect data in on-chip memory first, to perform efficient writes to global memory.
As every thread needs to know which row its entries belong to, we compute this assignment explicitly beforehand.
As the matrix is static during consecutive evaluations, we compute this information already in the build step.

\subsection{Fusion}
Kernel fusion is an important paradigm in parallel computing, to reduce  kernel launch overheads and costly memory transactions.
Whenever two operations in the high-level linear algebra formulation require the same input vectors and the number of threads required for both computations agree, the two kernels are merged, such that data does not have to be loaded multiple times.
For example in the crease module, the crease valency and vertex sharpness (Equations \ref{eq:CC_crease_valency} and \ref{eq:CC_crease_vsharpness} respectively) are computed.
Both computations involve the same left and right operands and only differ in the map, so these two mapped SpMVs are merged into a single kernel that computes both.
Similarly, when the output of a kernel is the input to the subsequent one, data does not have to go through global memory from the first to the second kernel but can directly be used if the kernels are merged.
If data causing the dependency is not used in any subsequent computation, the store to global memory is omitted.
Considering the crease example again, the crease valency computed in the first mapped SpMV is subsequently used as a map in the computation of the vertex sharpness.
As $C$ is symmetric we can parallelize over its columns $j$, compute $k_j$ and use it immediately afterwards to calculate $s_j$.

\begin{figure}
	\includegraphics[width=0.98\linewidth]{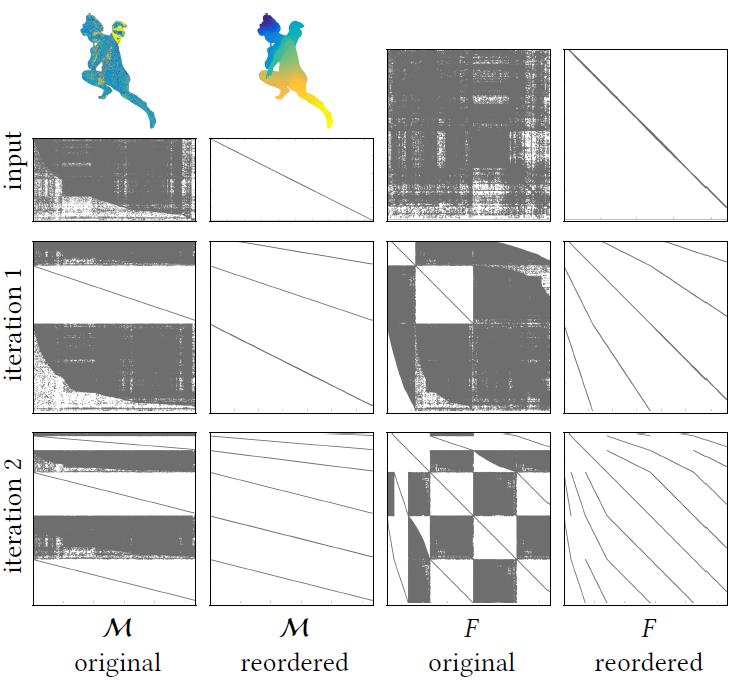}

	\caption{Evolution of the mesh matrix $\mathcal M$ of the original and RCM-reordered Angel model (first and second columns respectively) throughout two Catmull-Clark iterations. Color-coded geometric layouts of both orderings are shown on top. The evolution of the respective $F$ matrices is shown in the third and fourth columns.}
	\label{fig:matrixevol}
\end{figure}

\subsection{Mesh reordering}
Fast mesh querying is key to any efficient, high performance subdivision implementation.
While this may be sufficient in theory, non-algorithmic factors such as memory access and cache effects are crucial for algorithmic performance in practice.
Data layout in memory directly affects access patterns and therefore ensuring the locality of such patterns allows us to take advantage of caching mechanisms.
In this way, global reads and writes, which are known to cause performance deterioration, especially on GPUs, can be reduced.

In our context, this translates to ensuring that primitives which are topologically close in the mesh, reside close in memory as well.
The layout of a mesh in memory is reflected in the sparsity pattern of the mesh matrix, and locality can be enforced by clustering the non-zero elements close to the diagonal.
Closeness to the diagonal can be measured in terms of bandwidth~\cite{Davis2006}.
For bandwidth reduction, the reverse Cuthill McKee (RCM) algorithm is fairly well known to produce a comparatively inexpensive, low bandwidth reordering \cite{Cuthill1969,George1971}.
The original RCM only works on square symmetric matrices.
To reorder the (usually non-square) mesh matrices, we apply the RCM algorithm to the graph Laplacian of the mesh.
The acquired permutation is applied to the rows of  $\mathcal{M}$ and columns are sorted by their first non-zero entry.

As shown in Figure~\ref{fig:matrixevol}, there is no reason to expect mesh creators to deliver coherently ordered meshes.
Hence, a reordering of the input can be beneficial. 
Due to the way we build $M_{i+1}$, the mesh ordering deteriorates somewhat, as shown for the first iteration.
While the original vertex indices are copied to the next iteration, a \enquote{diagonal-like} line is added for the facepoints, and edge points are inserted depending on the connectivity of the original mesh.
If the original mesh followed a diagonal pattern, the edge points also correspond to a \enquote{diagonal-like} line.
Interestingly, the following subdivision iteration only adds two  \enquote{diagonal-like} lines for face and edge points.
Thus, all iterations show a somewhat tightly aligned sparsity pattern.
The same can be observed for $E$ and $F$, which both show a similar pattern, with $F$ being displayed in Figure~\ref{fig:matrixevol}.

%% file: results.tex
\section{Results}
\label{sec:results}

\begin{figure*}
    \includegraphics[width=.98\linewidth,valign=m]{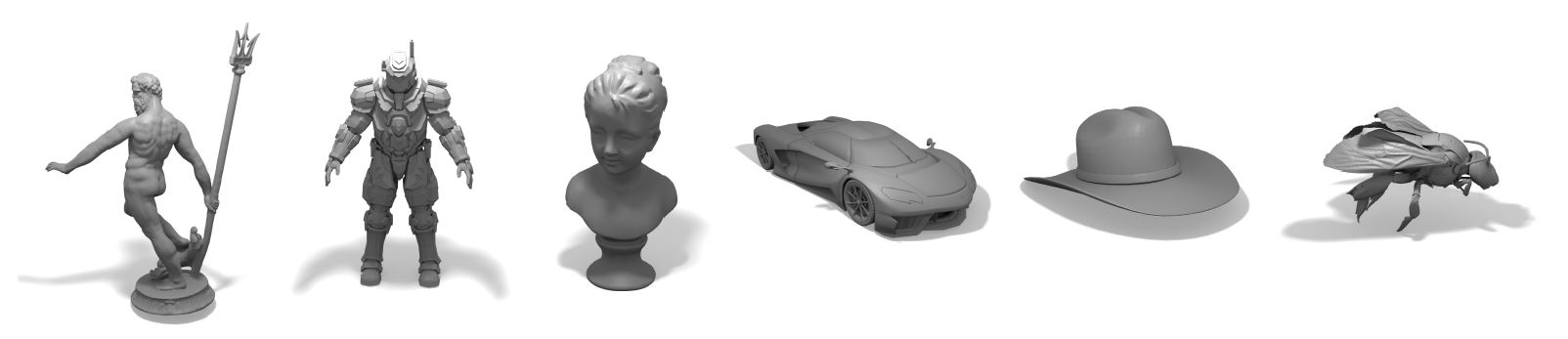}
	\caption{Selection of evaluation meshes: Neptune, ArmorGuy (courtesy of DigitalFish), Girl, Car, Hat, and  Eulaema Bee (courtesy of The Smithsonian).}
	\label{fig:tests2daltered}
\end{figure*}

We evaluate two variants of AlSub: formulating the algorithms in the language of sparse matrix algebra (\emph{AlSub pure}) as described in Section \ref{sec:catclark} and a version using the optimized kernels (\emph{AlSub opt.}) as described in Section \ref{sec:optimization}.
While there is a reasonable number of literature on parallel subdivision, there are hardly any implementations available for comparison.
Thus, we mainly compare to the current industry standard, OpenSubdiv, which is based on \cite{Niessner2012}.
OpenSubdiv splits subdivision into three steps.
First, a symbolic subdivision is performed to create refined topology, which is then used in a second step to precompute the stencil tables for vertex evaluation.
We summarize these two steps as \emph{build}.
The stencil tables are then used to perform the evaluation of refined vertex data (\emph{eval}).
While OpenSubdiv executes \emph{eval} on the GPU, \emph{build} runs entirely on the CPU.
To provide a comparison to a complete GPU implementation, we compare against \citeauthor{Patney2009}~\shortcite{Patney2009} .

All tests are performed on an Intel Core i7-7700 with 32GB of RAM and an Nvidia GTX 1080 Ti.
The provided measurements are the sum of all kernel timings required for the subdivision, averaged over several runs.
The input models are unaltered and thus have not been reordered unless specifically marked differently.

\subsection{Catmull-Clark performance}
As test models for  Catmull-Clark subdivision we use a variety of differently sized meshes which are listed in Table~\ref{tab:mesh_info_CC} with the tested subdivision level.
As AlSub is a modular approach which can adapt to different applications, we distinguish two use cases: \enquote{modeling} and \enquote{rendering}.

\begin{table}[h]
	\begin{tabular}{clrrrrr}
		\toprule
		                         & mesh      &  $c_f$ &  $c_v$ & $n_i$ & $r_f$ & $r_v$ \\ \midrule
		\multirow{13}{*}{\rtb{Catmull-Clark}}
								 & threeblock    &   18 &   20 &     5 & 18k  & 18k \\
								 & pig       &   381 &   389 &     5 & 390k  & 390k \\
								 & monsterfrog   &   1.3k &   1.3k &     4 & 331k  & 331k \\
								 & complex   &   1.4k &   1.3k &     4 & 346k  & 346k \\
								 & bigguy    &   1.5k &   1.5k &     4 & 371k  & 371k \\
		                         & ArmorGuy  &   8.6k &  10.0k &     6 & 35.2M & 35.3M \\
		                         & hat       &   4.4k &   4.4k &     6 & 18.1M & 18.1M \\
		                         & coat      &   5.6k &   5.7k &     6 & 22.8M & 22.8M \\
		                         & cupid     &    29k &   29k &      2 & 458k  & 458k \\
		                         & bike      &  53.9k &  54.3k &     4 & 13.4M & 13.5M \\
		                         & car       & 149.7k & 164.9k &     4 & 38.5M & 38.7M \\
		                         & dress     &   2.3k &   2.4k &     6 &  9.2M &  9.2M \\
		                         & bee       &  16.9M &   8.5M &     1 & 50.8M & 50.8M \\
		                         & neptune   &   4.0M &   2.0M &     2 & 48.1M & 48.1M \\
		                         \bottomrule
	\end{tabular}
	\caption{Catmull-Clark test meshes: Number of faces $c_f$ and vertices $c_v$ of the control meshes, as, well as the applied number of iterations $n_i$ and faces $r_f$ and vertices $r_v$ in the refined mesh.}
	\label{tab:mesh_info_CC}
\end{table}

\begin{figure*}
	\includegraphics[width=0.32\linewidth]{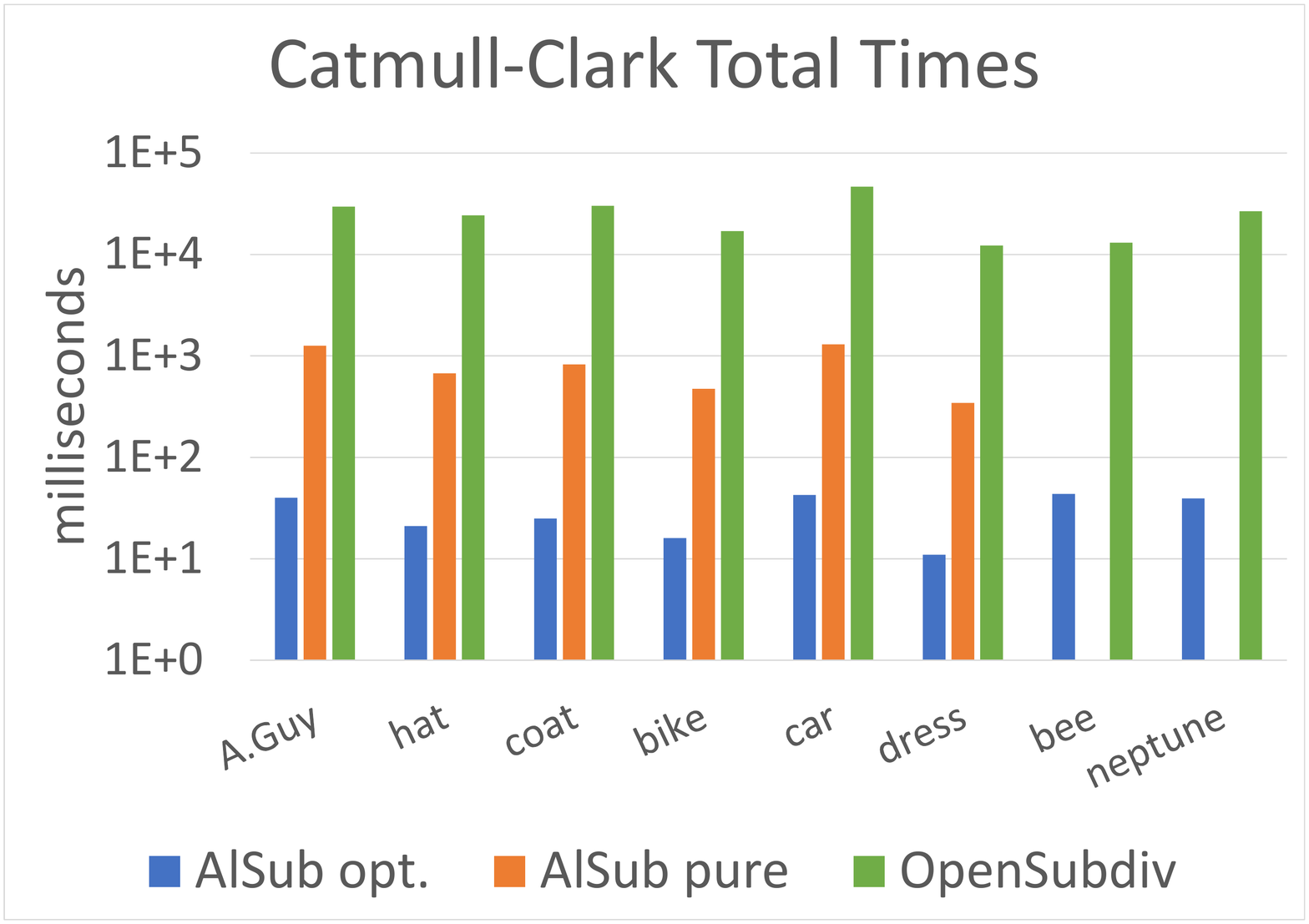}\hfill%
	\includegraphics[width=0.32\linewidth]{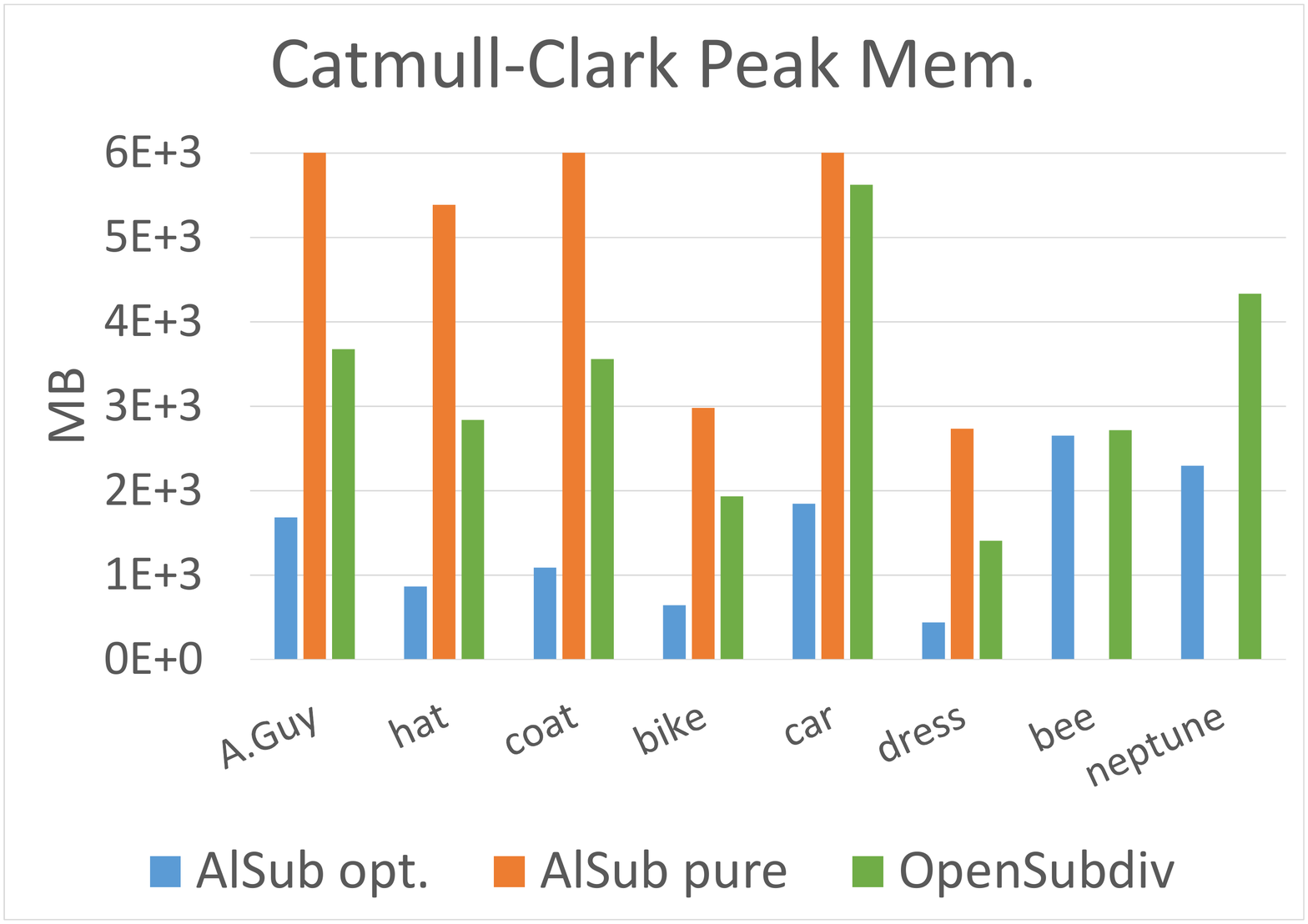}\hfill%
	\includegraphics[width=0.32\linewidth]{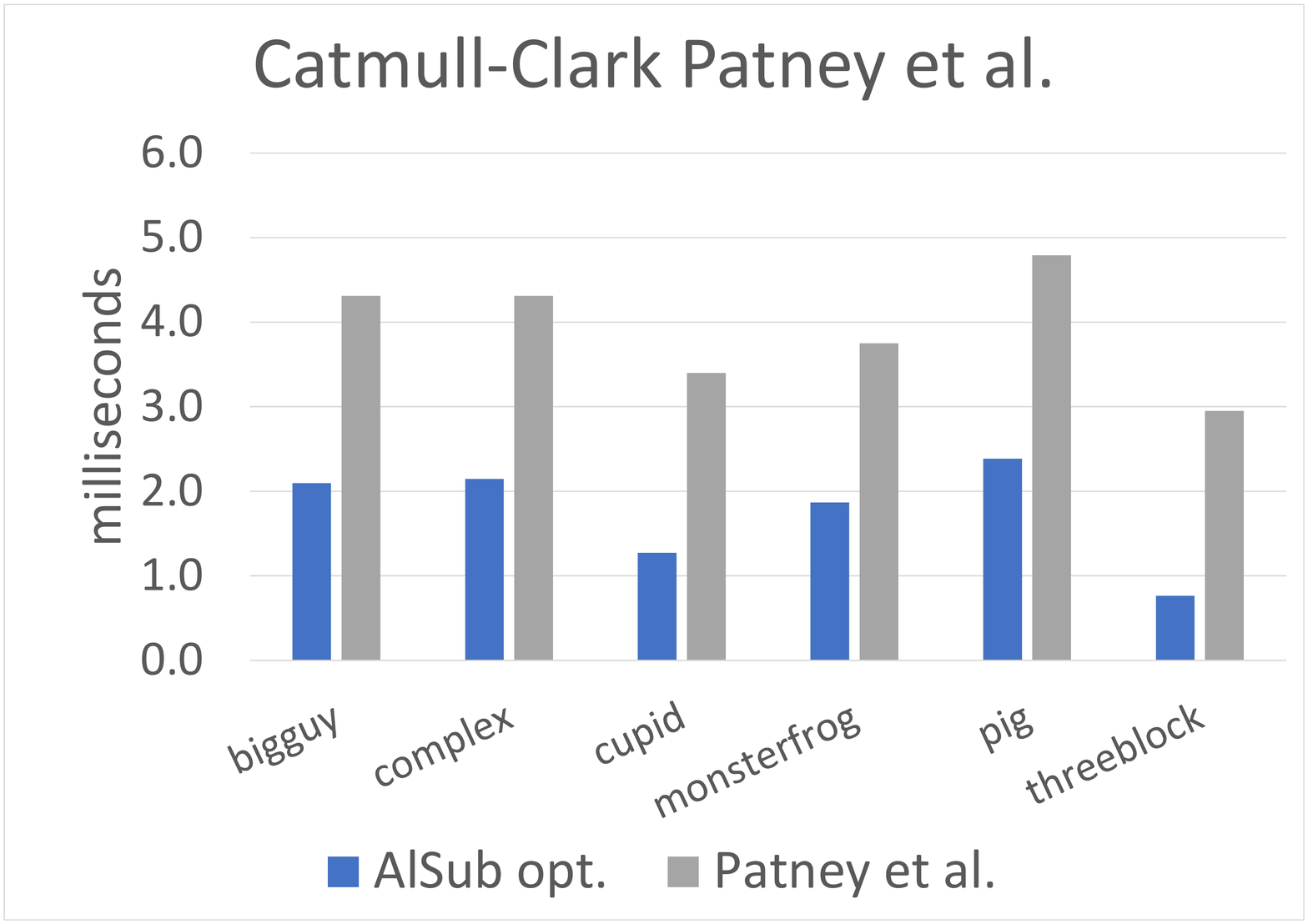}
	\caption{Catmull-Clark subdivision: Comparison of the complete subdivision time (topology + vertex positions) of AlSub and OpenSubdiv (left) and memory requirements (middle). A separate comparison for simpler meshes with the complete GPU-based approach by \citeauthor{Patney2009} (right). This is the time it takes after a topology changing modeling operation until the subdivided mesh is available.}
	\label{fig:res_modeling_CC}
\end{figure*}

\paragraph{Modeling}
This specific use case is a representative for the class of applications in which the mesh topology changes frequently.
Topological changes require re-computation of eventually preprocessed data \eg subdivision tables.
Results are given in Figure~\ref{fig:res_modeling_CC}.
The comparison between AlSub pure, AlSub opt., and OpenSubdiv (left) shows complex meshes.
We observe that AlSub opt. is more than one order of magnitude faster than AlSub pure, highlighting the gains of our optimizations.
AlSub pure is about one order of magnitude faster than OpenSubdiv when performing preprocessing and evaluation.
This shows that our complete parallel GPU implementation is significantly faster than the split CPU and GPU build and eval approach of OpenSubdiv if a topology-changing modeling operation is carried out.
Note that this is not the default use case of OpenSubdiv which assumes mostly static topology.
Nevertheless, this delay might still yield unpleasant behavior during topology changing modeling operations.
Note that the memory requirements for the entire subdivision of AlSub opt. are usually significantly below OpenSubdiv's stencil tables.
AlSub pure needs significantly more memory and thus was unable to execute on the very large meshes Bee and Neptune.

As OpenSubdiv is clearly more focused on efficient evaluation than optimizing the whole subdivision pipeline, we also compared to the GPU-based Catmull-Clark implementation by \citeauthor{Patney2009}~\cite{Patney2009}, which we configured to perform uniform subdivision.
We could only test small quad-only models, as their implementation seems to have issues when generating more geometry and fails on meshes with triangles.
Nevertheless, as seen in Figure~\ref{fig:res_modeling_CC} (right) AlSub opt. is about $2-3\times$ faster than the patch-based implementation of \citeauthor{Patney2009}.
We attribute this fact to the highly streamlined performance of our formulations and optimizations, which show zero redundant work and result in efficient memory movements.

\begin{figure}
	\centering
	\includegraphics[width=0.64\linewidth]{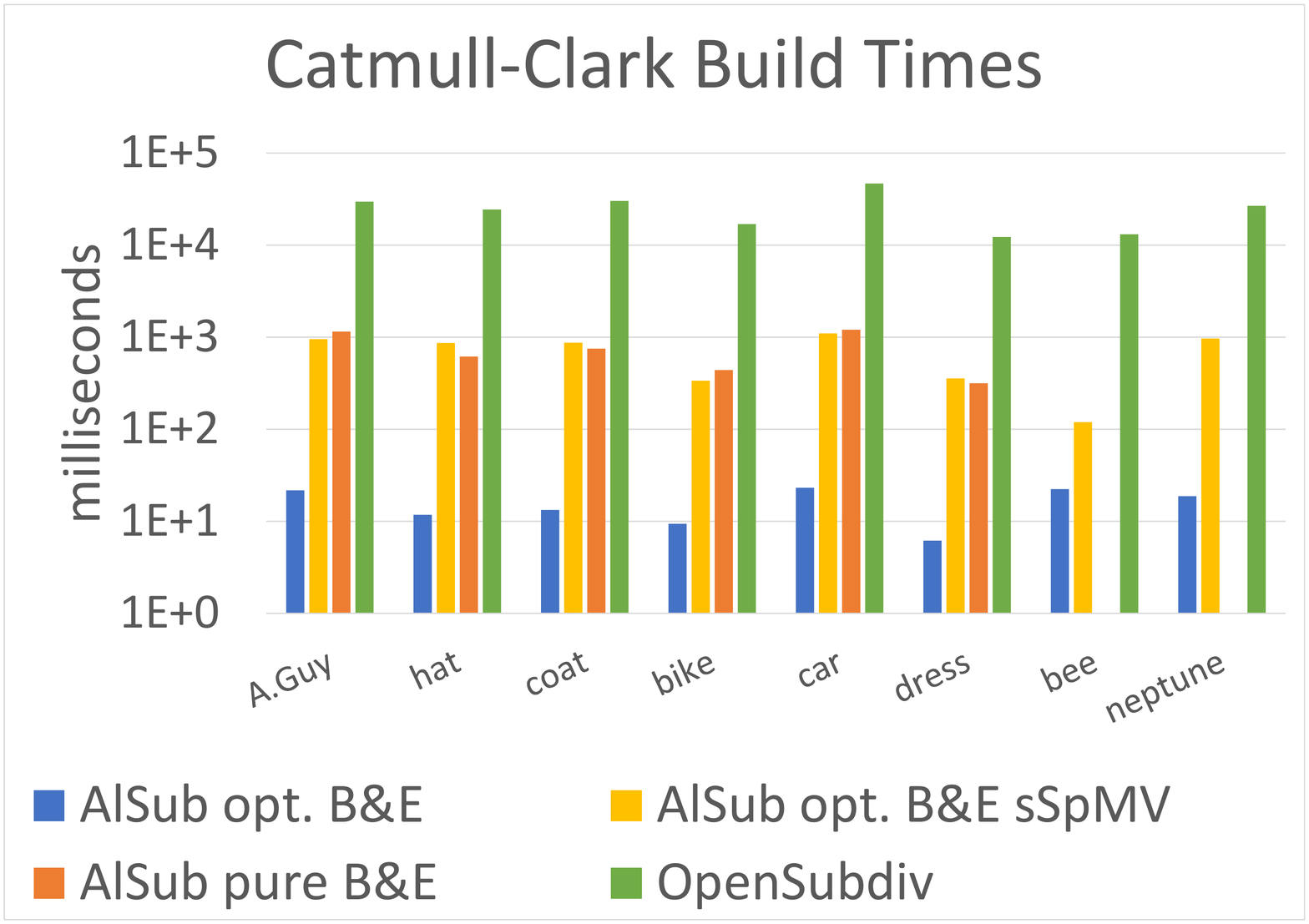}
	\includegraphics[width=0.64\linewidth]{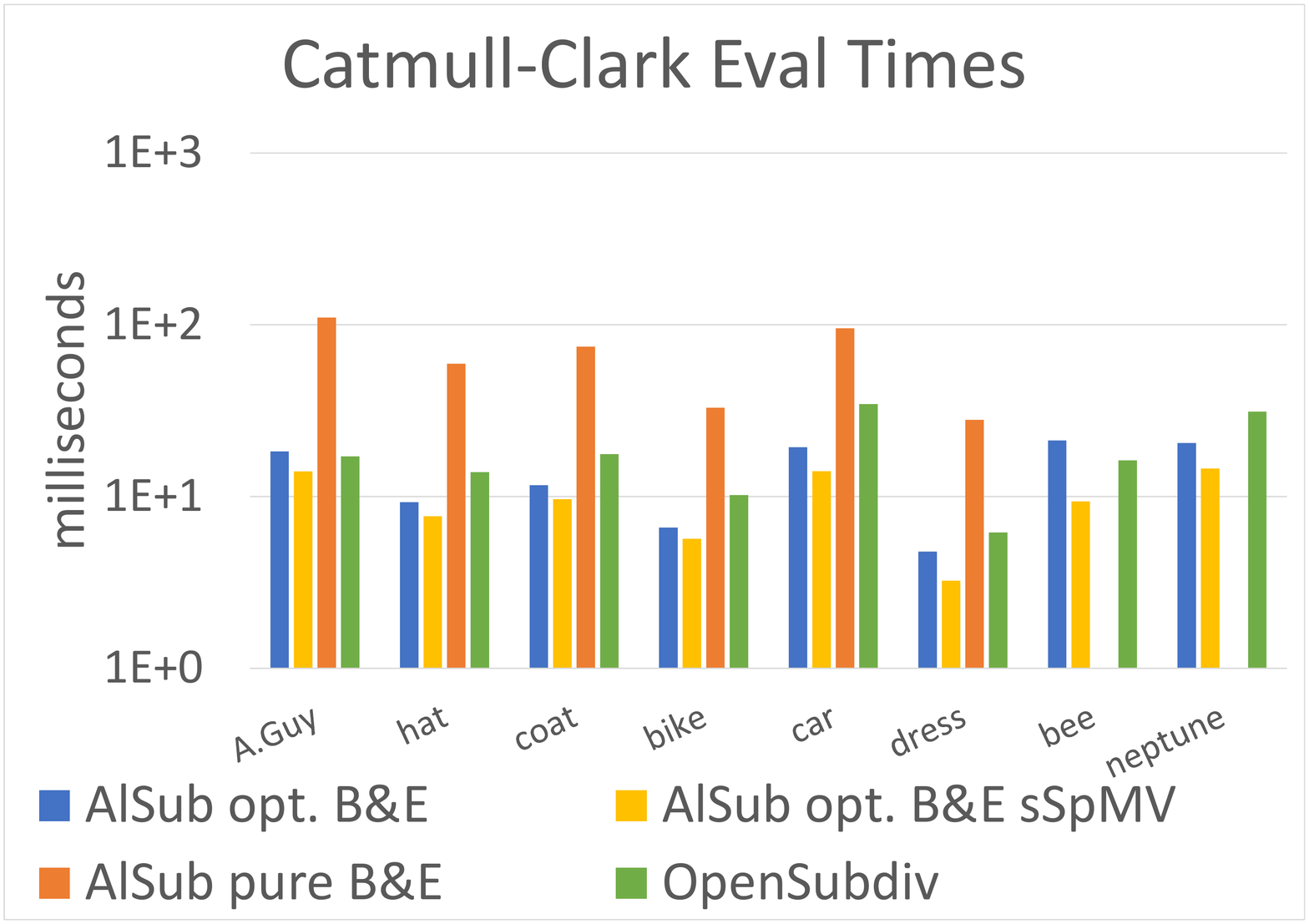}
	\caption{Catmull-Clark subdivision: Comparison of the individual steps (build \& eval) of AlSub and OpenSubdiv.}
	\label{fig:res_rendering_CC}
\end{figure}

\paragraph{Rendering}
In contrast to \emph{modeling}, topology is considered static in \emph{rendering}, which is the intended use case of OpenSubdiv.
In this case, information required during subdivision that only depends on the topology can be precomputed and stored for later use.
The evaluation stage uses this information to subdivide the vertex data in every render frame, \eg, when replaying an animation.
Relying on AlSub's split into \emph{build} and \emph{eval} modules, similar optimizations are possible in our approach.
Figure~\ref{fig:res_rendering_CC} shows that when splitting our approach into these two phases, AlSub opt. achieves nearly one order of magnitude performance gain over AlSub pure in both \emph{build} and \emph{eval}.
Furthermore, performing the \emph{build} stage complete on the GPU, yields significant performance gains over OpenSubdiv's build stage, even when building the matrix $\mathcal{R}$ for the single SpMV evaluation (AlSub opt. B\&E sSPMV).

Even more important for this use case is that AlSub opt. as well as AlSub opt. sSpMV also outperform OpenSubdiv in the \emph{eval} step. 
Comparing the customary AlSub opt. evaluation to the single SpMV module reveals a slight edge of sSPMV.
Performance of AlSub opt. decreases below that of OpenSubdiv for the Armor Guy model with its high number of creases, for which our approach in its current form performs additional steps to \enquote{fix} the geometry.
As creases are simply incorporated into the subdivision matrix $\mathcal{R}$ in the single SpMV evaluation, it does not suffer from that slowdown.
The good performance of AlSub's opt. eval is actually surprising, as it performs multiple mapped SpMVs for every subdivision level, while OpenSubdiv can perform the complete subdivison using a single kernel that only looks up the subdivision tables.
AlSub's opt. single SpMV evaluation leads the performance chart throughout all test cases.
However, it suffers from the same high memory usage as OpenSubdiv in the uniform case, as the subdivision matrix, as well as OpenSubdiv's stencil tables, might become prohibitively large. 
Performance for large models with a low subdivision depth (Bee and Neptune) is quite similar for all approaches.

\paragraph{Adaptive Subdivision}
To show that our approach can be used in a setting where only certain regions of the mesh have to be subdivided, we compare to the feature adaptive Catmull-Clark implementation of OpenSubdiv, which is based on the approach proposed by \citeauthor{Niessner2012}~\shortcite{Niessner2012}.
Here, only regions around irregularities have to be subdivided.
For regular mesh regions, patches are built which can be evaluated using hardware tessellation.
For evaluation, we use the Single SpMV variant of our approach.

\begin{figure*}
	\includegraphics[width=0.28\linewidth]{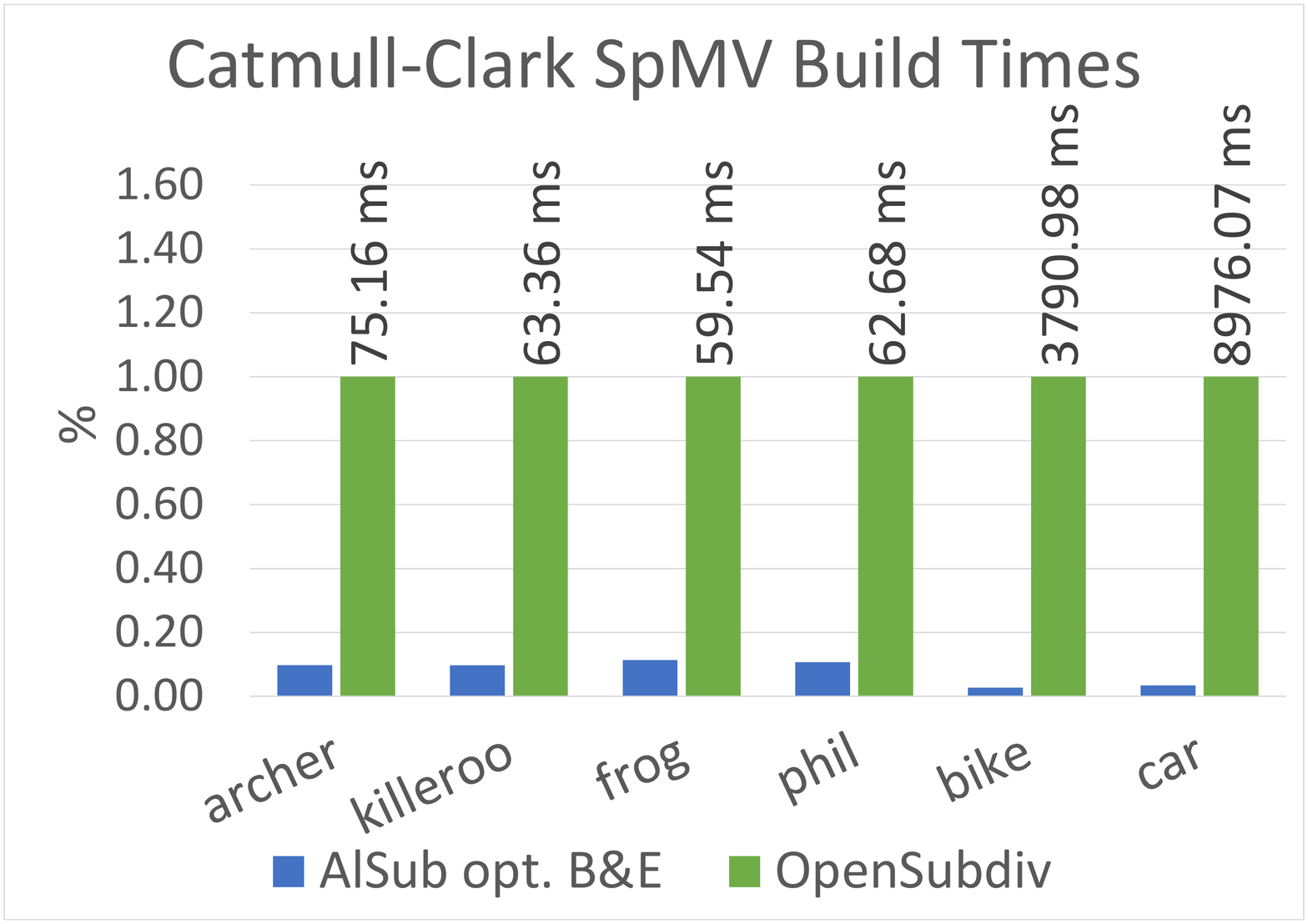}\hspace{20pt}
	\includegraphics[width=0.28\linewidth]{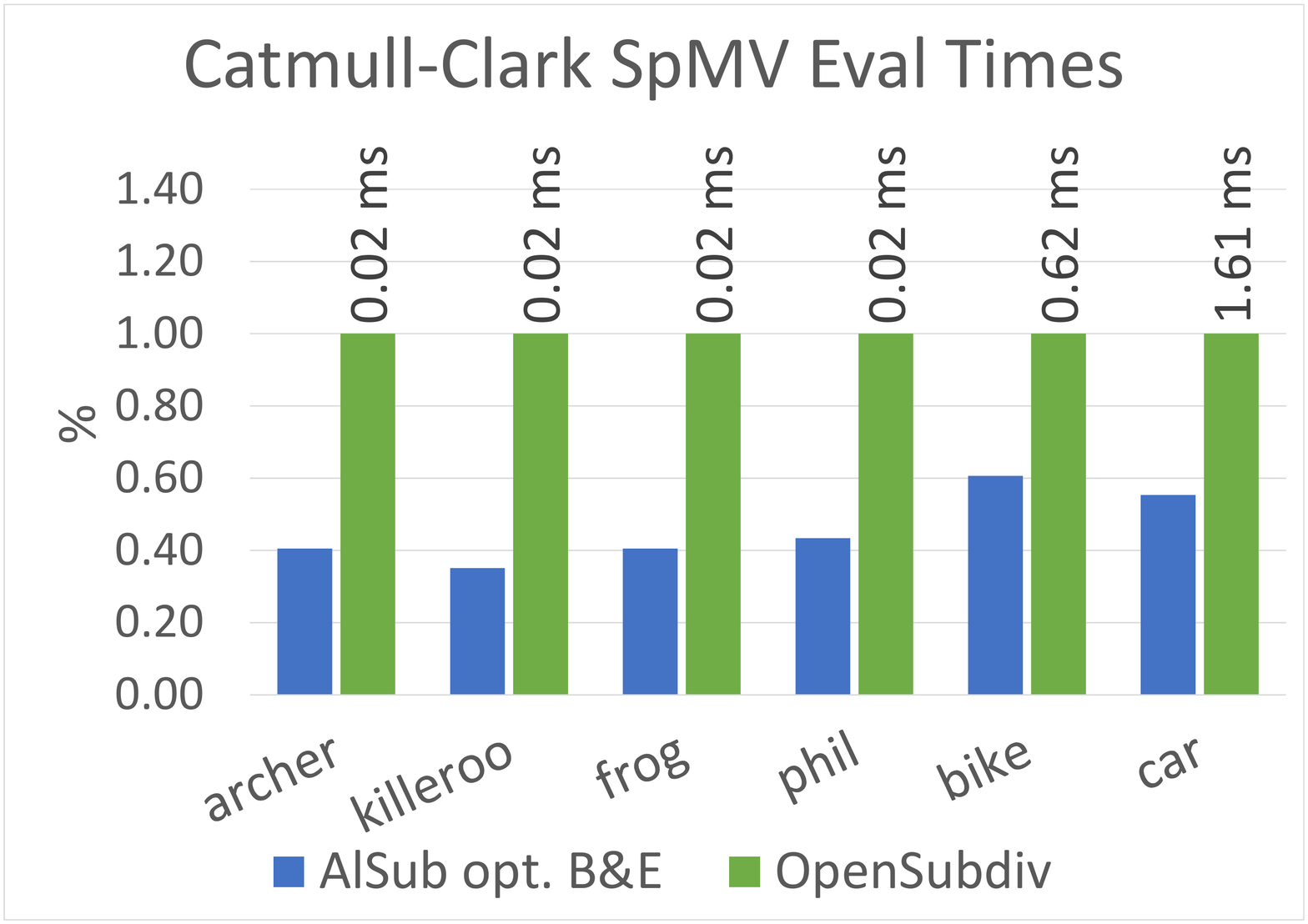}\hspace{20pt}
	\includegraphics[width=0.28\linewidth]{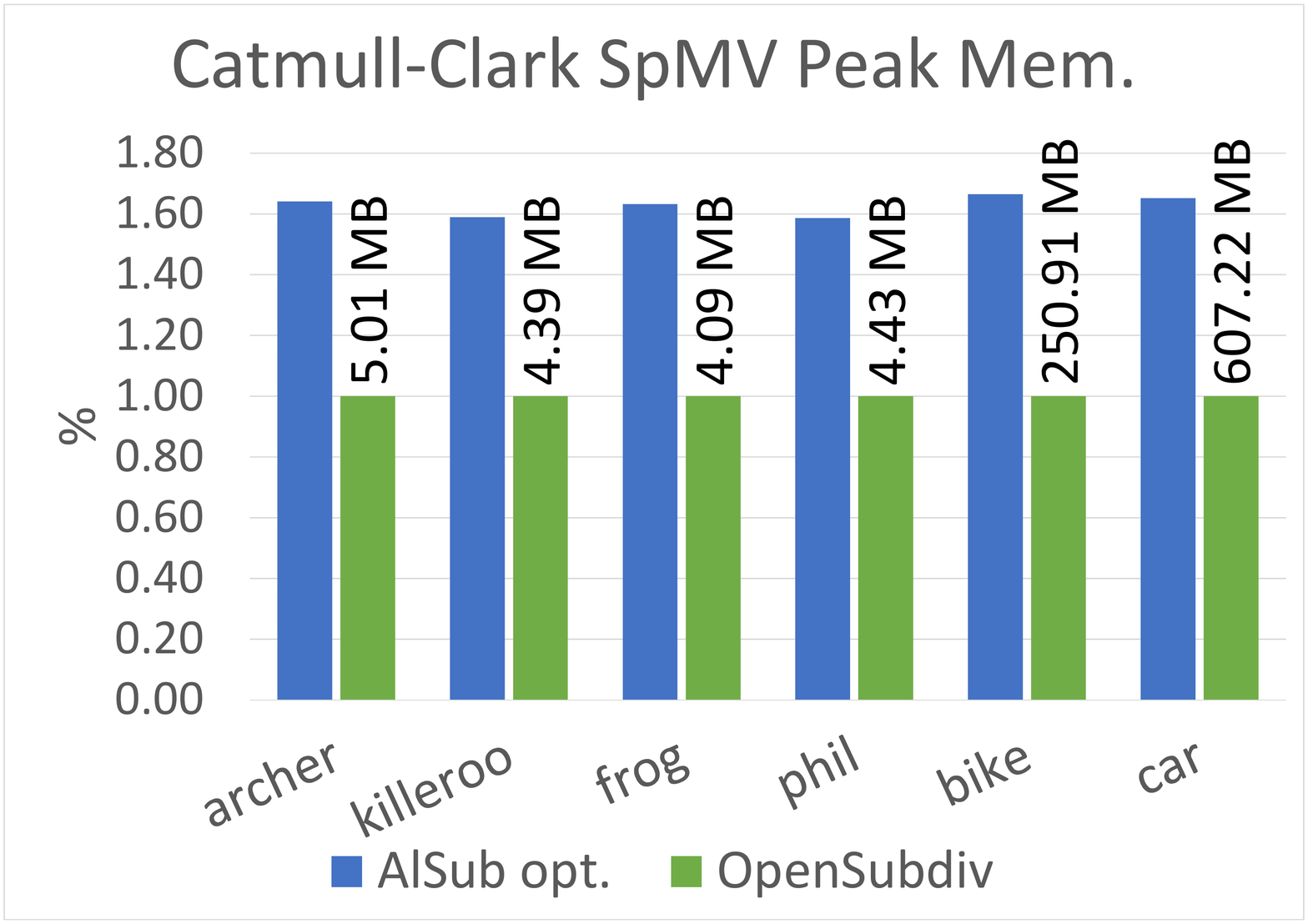}
	\caption{Adaptive Catmull-Clark: Comparison of the total time (build + eval) of AlSub opt. single SpMV relative to OpenSubdiv including peak memory consumption for both approaches.}
	\label{fig:res_CCA}
\end{figure*}

Figure \ref{fig:res_CCA} compares performance and required peak memory of our approaches with those of OpenSubdiv.
Again, performing the build step on the GPU shows a clear advantage of AlSub with a $17.7\times$ build speed-up.
The evaluation step is performed on the GPU in both approaches---while OpenSubdiv uses its stencil tables to create the refined vertex-data, AlSub performs a single SpMV.
While both approaches are similar in their nature, the simple optimizations applied to AlSub's SpMV evaluation reflect in an average performance increase of $2.3\times$ compared to OpenSubdiv's eval.
When comparing to the uniform subdivision from before, it can be observed that our optimizations work even better in this case with smaller $\mathcal{R}$ matrices.
We believe this is due to our load balancing strategies in the single SpMV which allows to draw more parallelism from the operations, which increases relative performance for small matrices. 
AlSub's peak GPU memory consumption is approximately $1.6\times$ higher than OpenSubdiv.
We attribute this to the fact that AlSub also does it's build step on the GPU---during which is reaches it's peak memory requirements---while OpenSubdiv only needs memory for the pre-computed stencil tables and vertex data.
The memory required during evaluation is similar in both approaches.

Considering the sum of all these results, AlSub seems to be a suitable drop-in replacement for OpenSubdiv in the modeling and rendering use case, virtually removing preprocessing costs and significantly increasing evaluation performance.

\begin{figure*}
	\includegraphics[width=0.32\linewidth]{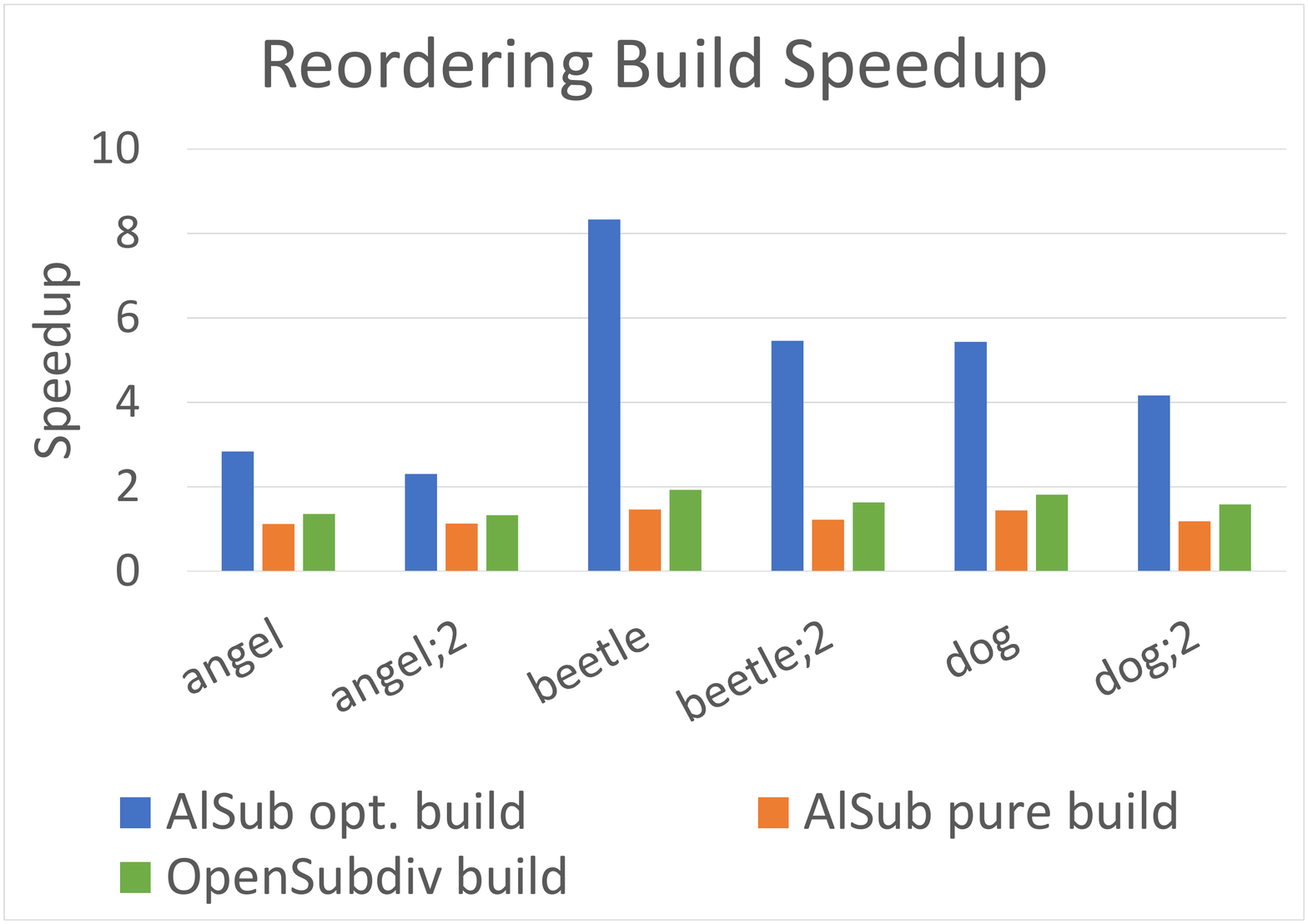}\hfill
	\includegraphics[width=0.32\linewidth]{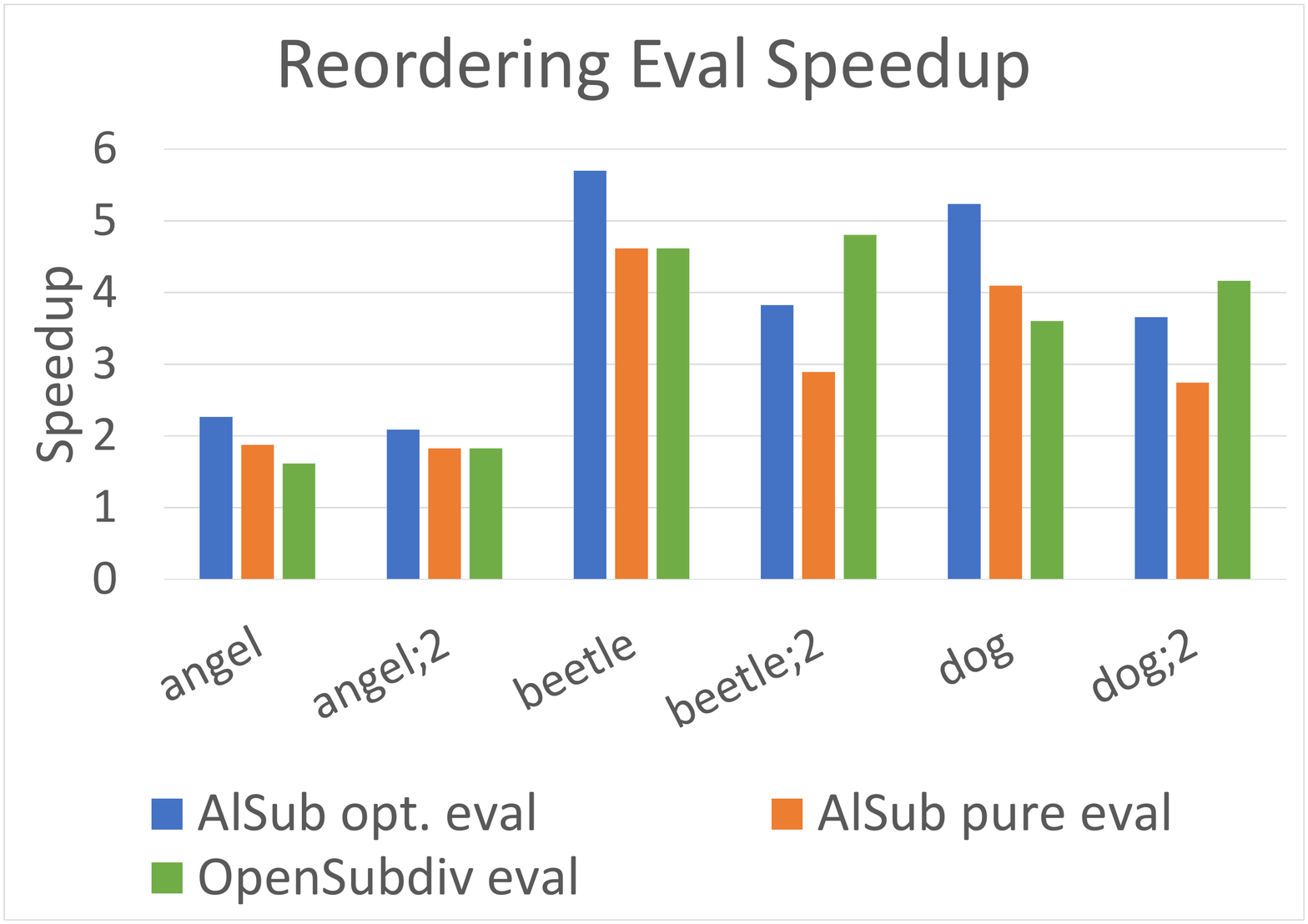}\hfill
	\includegraphics[width=0.32\linewidth]{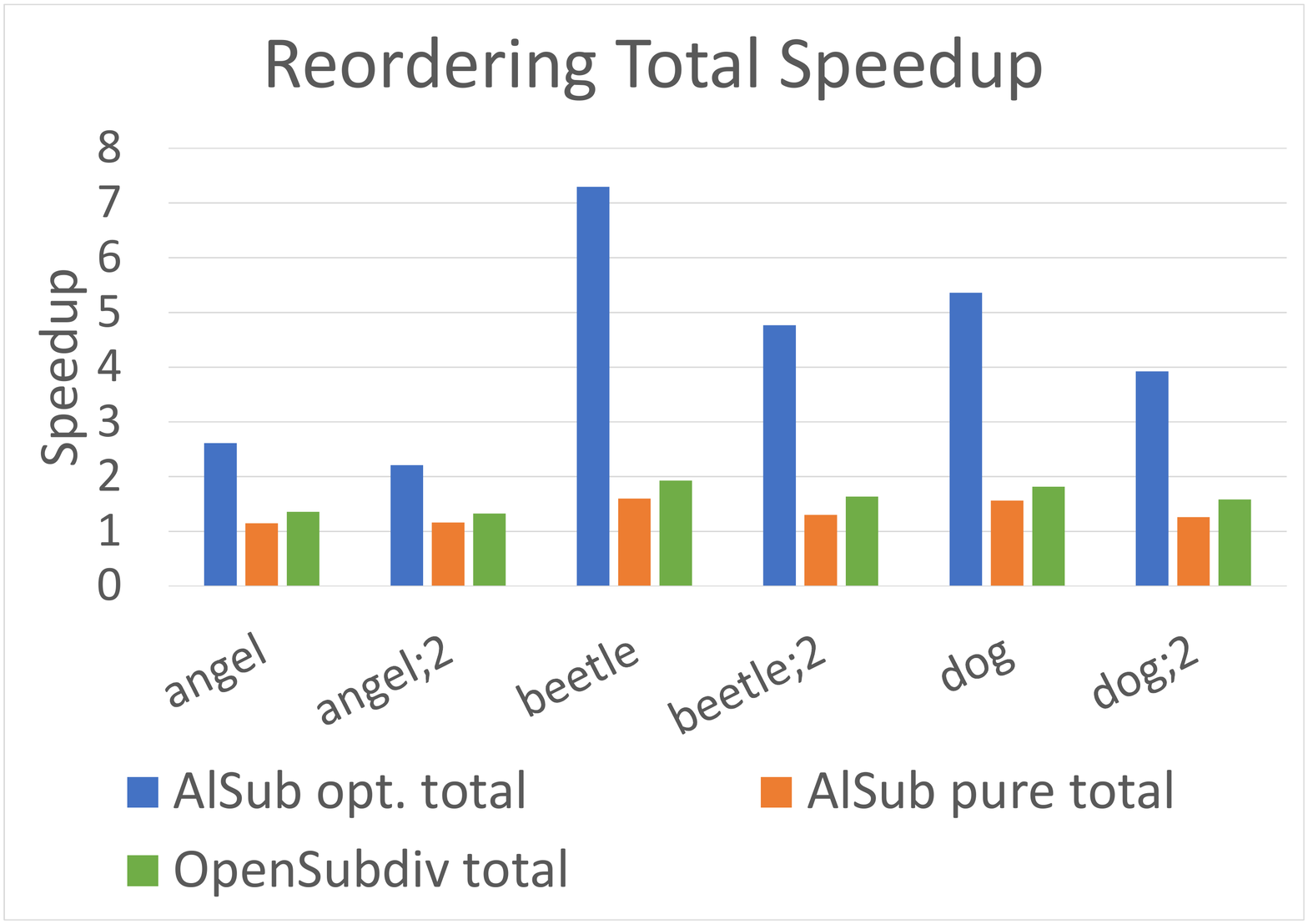}
	\caption{Reordering results as relative speed up to the non-reordered meshes over one Catmull-Clark iteration (no postfix) and two iterations (';2' postfix).}
	\label{fig:res_CC_re}
\end{figure*}

\paragraph{Mesh reordering}
To highlight the effect of mesh reordering, we compared the performance of different subdivision implementations on meshes in their original ordering and after reordering using the RCM method.
From the results in Figure~\ref{fig:res_CC_re}, it seems that our optimized kernels benefit most from better memory layouts.

Surprisingly, reordering can increase performance of up to $5-8\times$ in our optimized version for models which show a bad input data layout like the Beetle model.
While the speedups of other approaches are also significant, the relative speedup of AlSub opt. is on average slightly higher than for the other approaches.
We attribute the lower gains of SpLA to our optimized versions mainly operating on the input data directly due to kernel fusion while the SpLA version creates additional data structures, for which the data layout does not change significantly and always reduces performance compared to the optimized version.
OpenSubdiv gains its speedups mainly in the eval step, while AlSub improves performance equally among build and eval, indicating that the memory access pattern is more important on the GPU than the CPU.

Given the significant speedups which can be gained from reordering, it seems natural to attempt to find a fast reordering which can be used after each iteration to consolidate the memory layout. Our attempts in this direction suggest that it is a challenging problem since any gains get outweighed by the cost of reordering itself. Therefore, for scenarios such as production rendering where the topology does not change, it would be worthwhile to have reorderings precomputed for every few subdivision steps and deployed during batch processing tasks. 

\subsection{Loop and $\sqrt{3}$ performance}
To show that our approach is also efficient for other subdivision schemes, we show implementations for Loop and $\sqrt{3}$ subdivision.
The test meshes are outlined in Table~\ref{tab:mesh_info_L} and Table~\ref{tab:mesh_info_S}.

\begin{table}
		\begin{tabular}{clrrrlrr}
			\toprule
			& mesh      & $c_f$ & $c_v$ & $n_i$ & $r_f$ & $r_v$ &  \\ \midrule
			\multirow{8}{*}{\rtb{Loop}} & archer\_t &  3.2k &  1.6k &     6 & 13.1M &  6.5M &  \\
			& hat\_t    &  8.8k &  4.4k &     6 & 36.2M & 18.1M &  \\
			& goblet    &  1.0k &   520 &     6 & 4.1M  &  2.0M &  \\
			& Hhomer    & 10.2k &  5.1k &     6 & 41.8M & 20.9M &  \\
			& phil\_t   &  6.1k &  3.1k &     6 & 24.9M & 12.5M &  \\
			& star      & 10.4k &  5.2k &     6 & 42.5M & 21.3M &  \\
			& bee       & 16.9M &  8.5M &     1 & 67.8M & 33.9M &  \\
			& neptune   &  4.0M &  2.0M &     2 & 64.1M & 32.1M &  \\
			\bottomrule
		\end{tabular}
	\caption{Loop test meshes: Number of faces $c_f$ and vertices $c_v$ of the control meshes, as well as the applied number of iterations $n_i$ and faces $r_f$ and vertices $r_v$ in the refined mesh.}
	\label{tab:mesh_info_L}
\end{table}

\begin{table}
	\begin{tabular}{clrrrrr}
		\toprule
		& mesh       & $c_f$ & $c_v$ & $n_i$ &  $r_f$ &  $r_v$ \\ \midrule
		\multirow{7}{*}{\rtb{$\sqrt{3}$}}
		& fox        &   622 &   313 &     6 & 453.4k & 226.7k \\
		& girl\_bust & 61.3k & 30.7k &     6 &  44.7M &  22.4M \\
		& goblet     &  1.0k &   520 &     6 & 729.0k & 364.5k \\
		& Hhomer     & 10.2k &  5.1k &     6 &   7.4M &   3.7M \\
		& star       & 10.4k &  5.2k &     6 &   7.6M &   3.8M \\
		& bee        & 16.9M &  8.5M &     1 &  50.8M &  25.4M \\
		& neptune    &  4.0M &  2.0M &     2 &  36.1M &  18.0M \\
		\bottomrule
	\end{tabular}
	\caption{$\sqrt{3}$-subdivision test meshes: Number of faces $c_f$ and vertices $c_v$ of the control meshes as well as the applied number of iterations $n_i$ and faces $r_f$ and vertices $r_v$ in the refined mesh.}
	\label{tab:mesh_info_S}
\end{table}

\begin{figure*}
\begin{subfigure}{0.30\linewidth}
	\includegraphics[width=\linewidth]{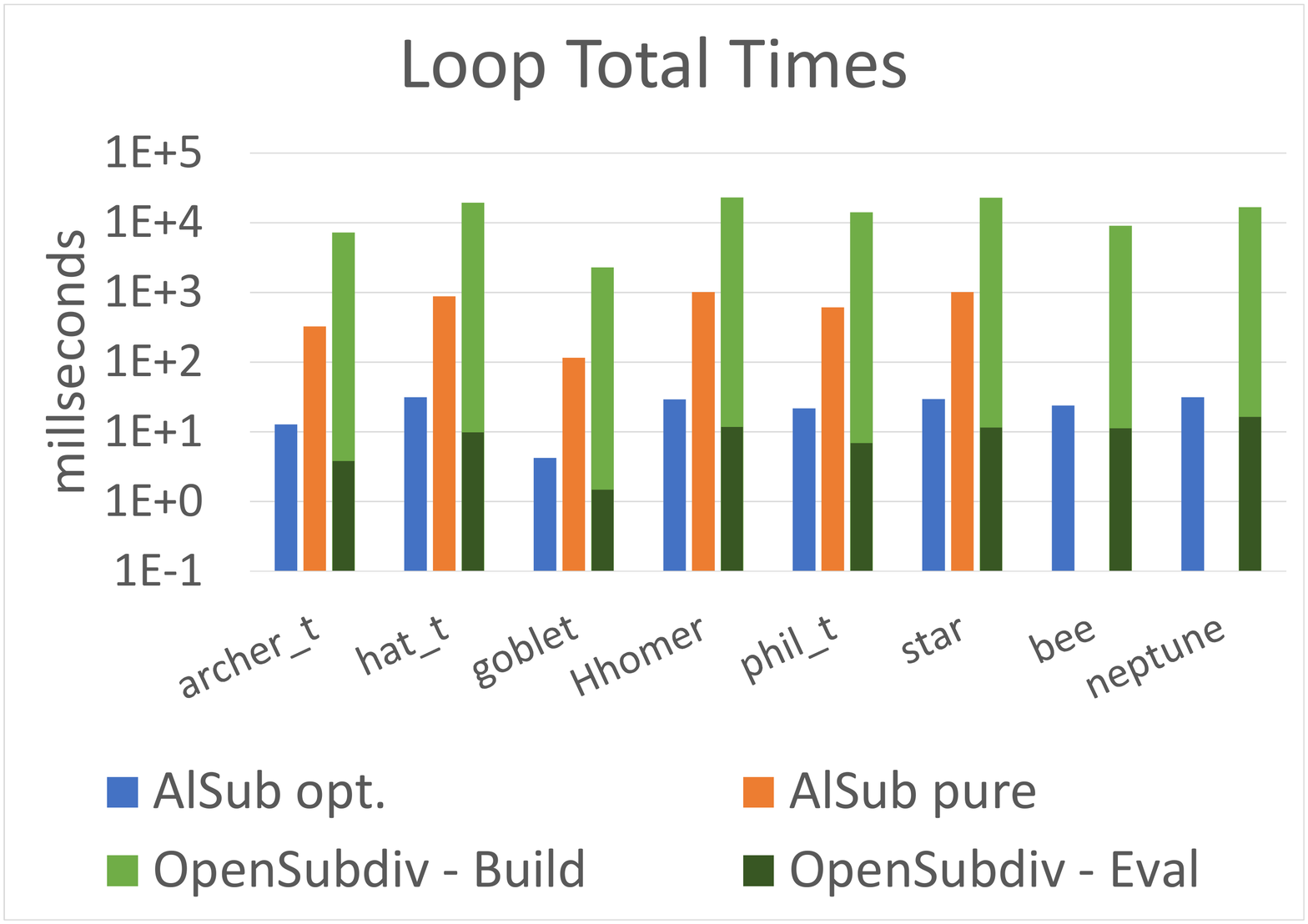}
	\caption{Loop subdivision performance}
	\label{fig:res_L}
\end{subfigure}\hfill%
\begin{subfigure}{0.30\linewidth}
	\includegraphics[width=\linewidth]{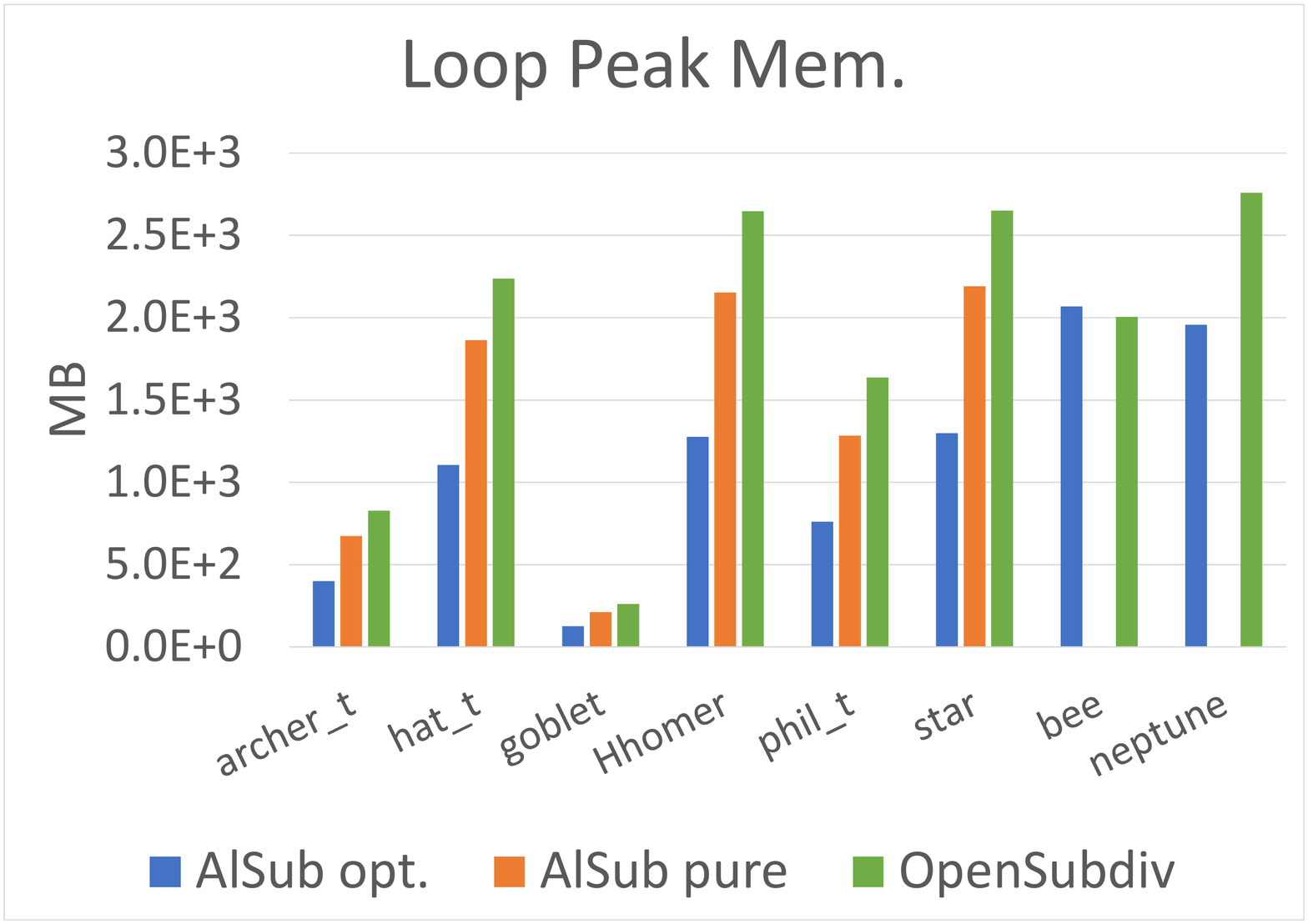}
	\caption{Loop memory requirements.}
	\label{fig:res_L_mem}
\end{subfigure}\hfill%
\begin{subfigure}{0.30\linewidth}
	\includegraphics[width=\linewidth]{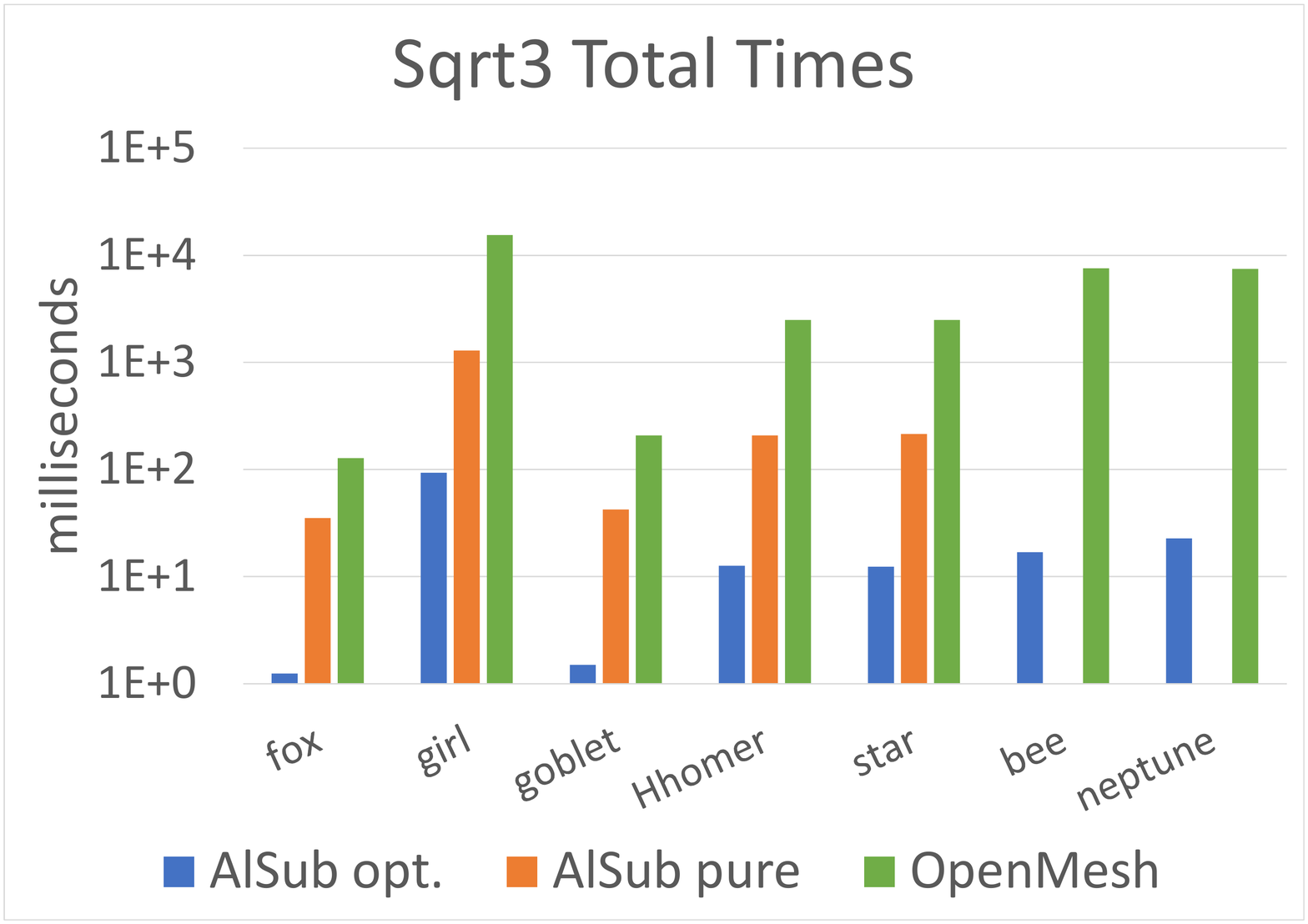}
	\caption{$\sqrt{3}$ performance}
	\label{fig:res_S}
\end{subfigure}
	\caption{Performance comparisons of AlSub's Loop and $\sqrt{3}$ subdivision implementations to OpenSubdiv's Loop and OpenMesh's $\sqrt{3}$.}
\end{figure*}

\paragraph*{Loop performance}
For Loop subdivison, we compare our approach to OpenSubdiv, as we were unable to find a more efficient comparison method.
Timing results are shown in Figure \ref{fig:res_L}.
Performance is similar to Catmull-Clark subdivision.
Again, our optimizations increase performance by about one order of magnitude.
Even when AlSub performs the full subdivision without any preprocessing, the execution times are similar to OpenSubdiv's eval step alone, which again is a single kernel mixing vertex weights.
The CPU build times of OpenSubdiv are significantly higher.
When splitting our approach into build and eval and performing single SpMV optimizations, similar speedups are achieved as before.
To reduce space, we omit that data here.
The memory requirements of AlSub opt. are again lower than AlSub pure. Interestingly, OpenSubdiv shows higher memory requirements than AlSub pure in many cases.

\paragraph*{$\sqrt{3}$ performance}
As OpenSubdiv lacks support for this scheme and we are not aware of any GPU implementation of $\sqrt{3}$, we compare AlSub pure, Alsub opt. and OpenMesh in Figure \ref{fig:res_S}.
While it is clear, that a parallel GPU implementation is capable of outperforming a serial CPU approach, it shows the benefits of a fully parallelized subdivision pipeline.
Throughout all experiments, the optimized AlSub opt. achieved a performance gain of $10\times$ or more compared to it's unoptimized counterpart.
Especially when starting with a smaller input model, AlSub opt. pulls away further, which we attribute mostly due the involved SpGEMM operations which show a certain overhead independent of the input size.
This overhead also reflects in the temporary memory requirements, which prohibit very large meshes (bee or neptune) to complete with our unoptimized version.
AlSub opt. handles these cases without trouble.

%% file: conclusion.tex
\section{Conclusion}
\label{sec:conclusion}
In this paper, we proposed a full fledged treatment of parallel mesh subdivision using linear algebra primitives (AlSub).
Our approach is modular and extensible, suitable for different subdivision schemes and handles additional features, like treatment of mesh boundaries, creases, and selective subdivision.
Unlike traditional approaches, where bookkeeping stalls performance and impedes vectorization, our treatment allows for efficient parallel implementations.
While a direct implementation of this formulation already indicates high throughput,
we show a series of optimizations that increase performance by another order of magnitude throughout all test cases.
The evaluation shows that our subdivision approach significantly outperforms other implementations in scenarios where an input mesh must be subdivided once ($2-3\times$ faster than patch-based GPU implementations).
Splitting the subdivision into preprocessing and evaluation for cases where the topology does not change, makes it a direct competitor to OpenSubdiv.
Performing the preprocessing step on the GPU shows extreme speedups compared to OpenSubdiv's CPU preprocessing.
At the same time, our evaluation step is up to $2\times$ faster than OpenSubdiv.
While we have shown how to write and optimize subdivision approaches in a complete sparse-linear algebra manner, many other types of mesh processing algorithms are equally suitable to this treatment.
We believe a general linear algebra library for mesh processing would benefit many domains.
Automatically identifying optimization potentials and applying similar steps as we have proposed in this work  may open up the high compute power of modern GPUs for many geometry processing algorithms. 

%% file: appendix.tex
\appendix

\section{$\sqrt{3}$-Subdivision}
\label{sec:sqrt3}

The $\sqrt{3}$-subdivision scheme is specialized for triangle meshes and is based on a uniform split operator which introduces a new vertex for every triangle of the input mesh~\cite{Kobbelt2000}. It defines a natural stationary subdivision scheme with stencils of minimum size and maximum symmetry. 

The subdivision process involves two major steps. The first one inserts a new vertex $f_i$ at the center of every triangle $i$.

\begin{figure}[!ht]
	\centering
	\includegraphics[width=.99\linewidth]{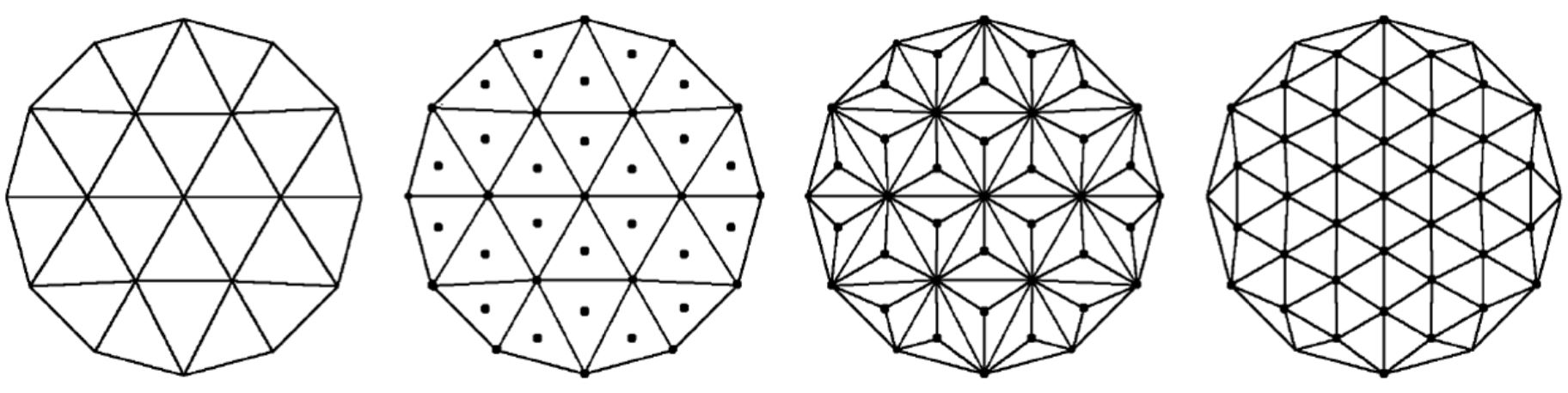}
	\caption{Original description of the $\sqrt{3}$-subdivision scheme. First a new vertex is inserted at every face of the given mesh. Second, an edge flip applied to the original mesh edges yields the final result, which is a $30$ degree rotated regular mesh. Applying this scheme twice leads to a $1$-to-$9$ refinement of the original mesh. Original image from ~\cite{Kobbelt2000}, copyright ACM.}\label{fig:subdiv1}
\end{figure}

Each new vertex is then connected to the vertices of its master triangle and an edge flip is then applied to the original edges, see Figure~\ref{fig:subdiv1}.
In the second step, the positions of the old vertices are updated using the following smoothing rule

\begin{equation}\label{equ:sqrt2}
S(p_i)=(1-\alpha_i) p_i +  \frac{\alpha_i}{n_i} \sum_1^{n_i} p_j
\end{equation}

where $n_i$ is the valence of vertex $p_i$ and $\alpha_n$ is obtained
by analyzing the eigen-structure of the subdivision matrix:
\begin{equation}\label{equ:alpha}
\alpha_{i}=\frac{4-2\cos(\frac{2\pi}{n_i})}{9}\text{.}
\end{equation}
Clearly the topological operations involved in this scheme anticipate an edge-based mesh representation and all the implementations we are aware of rely on the half-edge
data structure.

\begin{figure}[!h]
	\centering
	\includegraphics[clip, trim =0cm 6cm 0cm 6cm, width=0.48\textwidth]{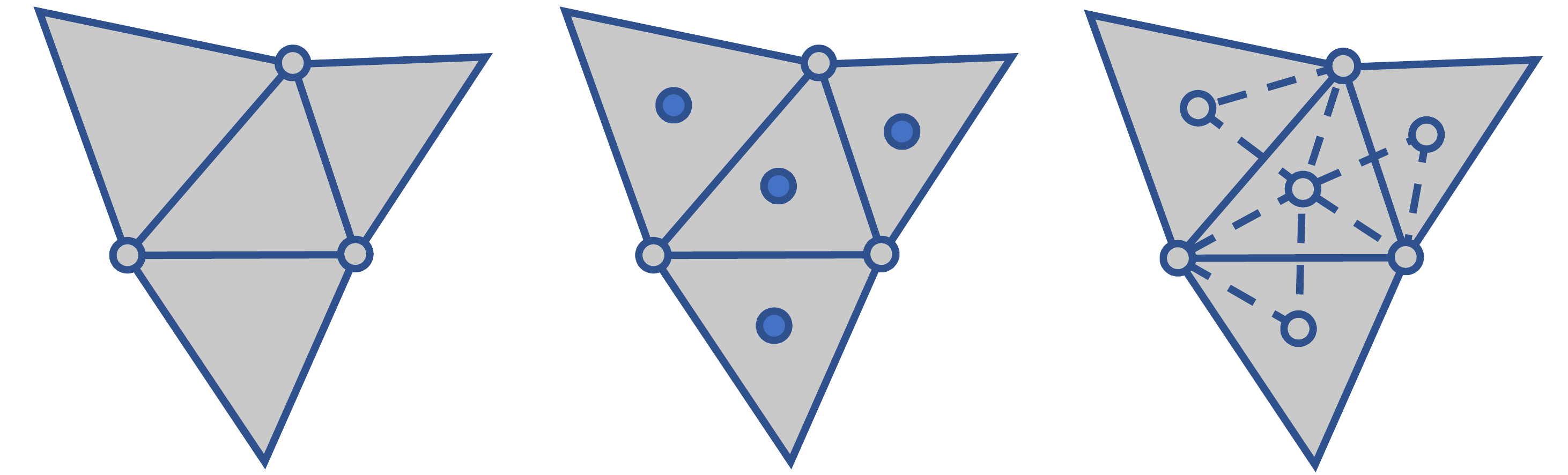}
	\caption{ After inserting the new vertices (blue), each triangle contributes three new triangles to the refined mesh which are obtained by joining its vertices to their respective right and left neighboring new vertices.}\label{fig:subdiv2}
\end{figure}

In order to adapt this subdivision scheme to our matrix algebra framework, we reinterpret the whole process in a slightly different manner. By reasoning only on triangles as detailed in Figure~\ref{fig:subdiv2}, the topological operations get simplified and the subdivision scheme can be easily abstracted using sparse matrix algebra. In fact, we need only a good bookkeeping of triangle-triangle adjacency to obtain new triangulations and update vertex positions.
Please note that the boundary can be treated by using adequate smoothing ~\cite{Kobbelt2000} using similar ideas to the outline given earlier for the Catmull-Clark scheme, but we omit it here to keep the presentation succinct.

\paragraph{New vertex points}
A new vertex is added to each triangle's barycenter.
The average of triangle vertices can be calculated using the mapped SpMV
\begin{equation}
\mathbf{f} = \underset{(1,2,3)\rightarrow\frac{1}{3}}{\mathcal{M}^T\mathbf{P}}
\end{equation}

\paragraph{Required adjacency information}
The $\sqrt{3}$ scheme adds a vertex to each triangle and connects it to the new vertices on the three neighboring triangles.
To find these neighbors efficiently, we can again use the oriented graph adjacency matrix to store the index of the adjacent face to any given edge, as in Equation \ref{eq:F}.

\paragraph{Vertex update}
The second term in the smoothing step can also be performed with action maps
\begin{equation}
\underset{ val_i \rightarrow \frac{\alpha_i}{n_i}}{F \mathbf{P}}
\end{equation}

\paragraph{Topology refinement}
For a mesh with $n_v$ vertices, each vertex of a given triangle $(p_k,p_l,p_m)$ with index $i$ contributes a new triangle to the refined mesh. For instance, vertex $p_k$ contributes the triangle consisting of $p_k$ itself, the face-point $f_i$ which can be conveniently indexed by $i + \lvert v \rvert$ and the barycenter on an adjacent triangle, which then takes index $F(l,k) + \lvert v \rvert$.
The mesh matrix of the refined mesh can be efficiently created in parallel by appending new columns.

\section{Loop Subdivision}
\label{sec:loop}

\begin{figure}[ht]
	\includegraphics[clip, trim =0cm 2cm 0cm 2cm, width=0.48\textwidth]{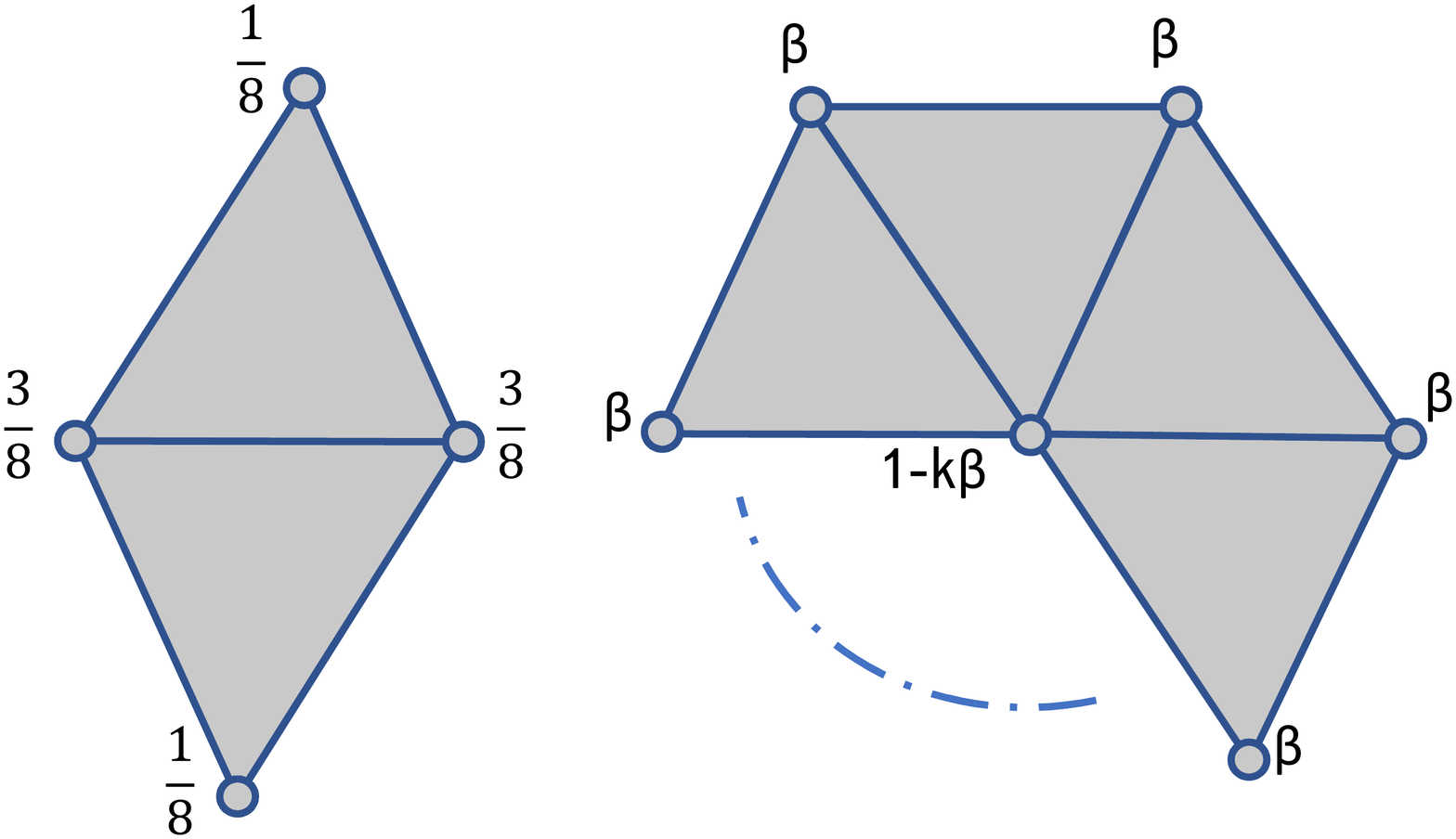}
	\caption{For each edge, the Loop scheme inserts a new vertex as a weighted sum of the vertices of the adjacent triangles (left). In the smoothing step, original positions are updated using a $\beta$-weighted combination of the their neighbors (right).}
	\label{fig:loop_scheme}
\end{figure}

This scheme is another triangle mesh subdivision method which was introduced by \citeauthor{Loop1987}~\shortcite{Loop1987}.
It refines a mesh by inserting new edge-points as described in Figure~\ref{fig:loop_scheme}-left.
For each triangle, these points can be used to perform a split into four new triangles.
The original vertex positions are then smoothed using local weighted averaging as summarized in Figure \ref{fig:loop_scheme}-right.
The weighted average in the smoothing step is based on convergence consideration and is defined as
\begin{equation}\label{eq:loop_smooth}
S(p_i)=(1-n_i\beta_i) p_i +  \beta_i \sum_1^{n_i} p_j,
\end{equation}
where
\begin{equation}
\beta_{i} = \frac{1}{n_i}\cdot\left(\frac{5}{8}-\left(\frac{3}{8} + \frac{1}{4} \cdot \cos\left(\frac{2\pi}{n_i}\right)\right)^2\right).
\end{equation}

In the following, we briefly describe the algebraic machinery we use to capture the topological modifications intrinsic to this scheme.
\paragraph{New vertex points}
For a given edge, the vertex insertion requires the edge vertices and the vertices opposite to the edge. We can gather this information by using the adjacency matrix of the directed graph of the mesh and for each edge store the index of the remaining triangle vertex as the non-zero value.

\begin{equation}
G = \underset{\lbrace Q_{3}\rbrace  \lbrack \lambda \rbrack}{\mathcal{M}\mathcal{M}^T}
\end{equation}
\begin{equation}
\lambda(i,j) =
\begin{cases}	
k & if \quad Q_3 = 1 \\	
0 & else	
\end{cases}
\end{equation}

where $k$ is the vertex opposite to edge $p_ip_j$.

Unique edge indices can be obtained from this matrix by summing it with its transpose to obtain a matrix $E$ and incrementally assigning indices to the non-zeros of the upper triangular part of $E$ similarly to how it is done in the context of Catmull-Clark subdivision.
The new vertex locations can then be obtained by looking up the unique edge indices and for each edge $p_ip_j$,  obtaining the opposite vertices as $G(i,j)$ and $G(j,i)$ and performing the summation as given in Figure~\ref{fig:loop_scheme}-left.

\paragraph{Vertex update}
The second term in Equation~\ref{eq:loop_smooth} can be computed using the mapped SpMV below
\begin{equation}
\underset{ val_i \rightarrow \beta_i}{G \mathbf P},
\end{equation}
where the action maps substitutes values in row $i$ by $\beta_i$.

\paragraph{Topology refinement}
For each triangle $(p_k,p_l,p_m)$ in the control mesh, three new triangles of the refined face are simple arrangements of an original vertex and two new edge-points.
The fourth triangle is only composed of the three new edge-points.
The original vertices' indices are $p_k$, $p_l$, $p_m$ and unique edge indices can be obtained from the upper triangular part of $E$, as done in the Catmull-Clark scheme.
With this information, the refined mesh matrix can be constructed efficiently in parallel.